\documentclass[letterpaper,twocolumn,10pt]{article}
\usepackage{usenix-2020-09}
\usepackage{url}

\AtBeginDocument{%
	\providecommand\BibTeX{{%
			\normalfont B\kern-0.5em{\scshape i\kern-0.25em b}\kern-0.8em\TeX}}}

\usepackage{setspace}
\usepackage{longtable}
\usepackage[figuresright]{rotating}
\usepackage{epsfig,endnotes,color}
\usepackage{graphicx}
\usepackage{amsmath}
\usepackage{caption}
\usepackage{graphicx}
\usepackage{url}
\usepackage{xcolor}
\usepackage{float}
\usepackage{lipsum}
\usepackage{listings}
\definecolor{codegreen}{rgb}{0,0.6,0}
\definecolor{codegray}{rgb}{0.5,0.5,0.5}
\definecolor{codepurple}{rgb}{0.58,0,0.82}
\definecolor{backcolour}{rgb}{0.95,0.95,0.95}

\usepackage{svg}
\usepackage{CJKutf8}
\usepackage{subfigure}
\usepackage{lscape, latexsym, amssymb, algorithmic, multirow}
\usepackage[linesnumbered, vlined, ruled]{algorithm2e}
\usepackage{multirow}
\usepackage{mathtools, bbm, color}
\usepackage{booktabs}
\usepackage{amsthm,mathrsfs,amsfonts,dsfont}
\usepackage{enumerate}
\usepackage{hhline}
\usepackage{enumitem}
\usepackage{tabu}
\usepackage{colortbl}
\usepackage{enumerate}
\usepackage[graphicx]{realboxes}

\usepackage{footnote}
\usepackage{color}
\usepackage{colortbl}
\usepackage{pifont}
\usepackage{bbding}
\usepackage{threeparttable}
\usepackage{rotating}
\usepackage{authblk}

\def\BibTeX{{\rm B\kern-.05em{\sc i\kern-.025em b}\kern-.08em
    T\kern-.1667em\lower.7ex\hbox{E}\kern-.125emX}}

\title{\system: Detecting Third-Party Component Usage Violations in IoT Firmware}

\newcommand{\system}{{\sc UVscan}\xspace}

\newcommand\blfootnote[1]{%
\begingroup
\renewcommand\thefootnote{}\footnote{#1}%
\addtocounter{footnote}{-1}%
\endgroup
}

\newfont{\mycrnotice}{ptmr8t at 7pt}

\begin{document}
\author[$^\ast$, $^\dag$]{Binbin Zhao}
\author[$\dag$]{Shouling Ji}
\author[$\dag$]{Xuhong Zhang}
\author[$\ddag$]{Yuan Tian}
\author[$\dag$]{Qinying Wang}
\author[$\dag$]{Yuwen Pu}
\author[$\dag$]{Chenyang Lyu}
\author[$^\ast$]{Raheem Beyah}

\affil[ ]{
$^\ast${\small Georgia Institute of Technology},
$^\dag${\small Zhejiang University},
$^\ddag${\small  University of California, Los Angeles}
}

\affil[ ]{{\small E-mails: binbin.zhao@gatech.edu, 
\{sji, zhangxuhong\}@zju.edu.cn, yuant@ucla.edu, \{wangqinying, yw.pu, puppet\}@zju.edu.cn, rbeyah@coe.gatech.edu.}}

\maketitle
\begin{abstract} 
Nowadays, IoT devices integrate a wealth of third-party components (TPCs) in firmware to shorten the development cycle. 
TPCs usually have strict usage specifications, e.g., checking the return value of the function. 
Violating the usage specifications of TPCs can cause serious consequences, e.g., NULL pointer dereference. Therefore, this massive amount of TPC integrations, if not properly implemented, will lead to pervasive vulnerabilities in IoT devices.  
Detecting vulnerabilities automatically in TPC integration is challenging from two perspectives:
(1) There is a gap between the high-level specifications from TPC documents, and the low-level implementations in the IoT firmware. (2) IoT firmware images are mostly closed-source binaries, which lose lots of information when compiling from the source and have diverse architectures.

To address these challenges, we design and implement \system,  an automated and scalable system to detect TPC usage violations in IoT firmware. In \system, we first propose a novel natural language processing (NLP)-based rule extraction framework, which extracts API specifications from inconsistently formatted TPC documents. We then design a rule-driven NLP-guided binary analysis engine, which maps the logical information from the high-level TPC document to the low-level binary, and detects TPC usage violations in IoT firmware across different architectures. We evaluate \system from four perspectives on four popular TPCs  and six ground-truth datasets.
The results show that \system achieves more than 70\% precision and recall, and has a significant performance improvement compared with even the source-level API misuse detectors. To provide an in-depth status quo understanding of the TPC usage violation problem in IoT firmware, we conduct a large-scale analysis on 4,545 firmware images and detect 27,621 usage violations. Our further case studies, the Denial-of-Service
attack and the Man-In-The-Middle attack on several firmware
images, demonstrate the serious risks of TPC usage
violations. Currently, 206 usage violations have been confirmed by vendors as vulnerabilities, and seven of them have been assigned CVE IDs with high severity. 
\end{abstract}
\blfootnote{Shouling Ji is the corresponding author.}

\section{Introduction}
Nowadays, the Internet of Things (IoT) is omnipresent and plays an essential role in our daily lives. According to a recent report~\cite{dataprot}, 152,200 IoT devices will be connected to the internet per minute by 2025 and the number of active IoT devices will exceed 25.4 billion by 2030. Nevertheless, the growing number of IoT devices poses many serious security risks. For instance, a large number of IoT devices, e.g., routers and IoT gateways, are vulnerable to the NULL pointer dereference vulnerability due to the API misuses in the closed-source Qualcomm QCMAP suite~\cite{qcmapvul}, making them ideal compromise targets of various botnets, e.g., Mirai~\cite{mirai2017usenix, zhao2020large}.



Currently, the core part of IoT devices, \emph{firmware}, widely integrates a large number of third-party components (TPCs) to facilitate development efficiency~\cite{zhao2022large, zhao2023tdsc}.  A TPC always comes with a lot of built-in functions, also referred to as application programming interfaces (APIs). These APIs tend to have complicated specifications, which are typically written with lengthy statements in TPC documents. Nevertheless, developers may fail to strictly follow the API specifications when adopting TPCs, resulting in the \emph{TPC usage violation problem}, which is typically caused by a set of API misuses, e.g., unchecked return value and incorrect invocation sequence. Many previous works have indicated that the API misuse will introduce serious security implications~\cite{DBLP:conf/icse/PanditaXZXOP12, DBLP:conf/icsm/PanditaTWT16,DBLP:conf/uss/YunMSJKN16, DBLP:conf/kbse/KangRJ16, lv2020rtfm, shoshitaishvili2015firmalice}. For instance, the incorrect use of OpenSSL APIs can cause  Man-In-The-Middle  attacks  ~\cite{DBLP:conf/ccs/GeorgievIJABS12, DBLP:conf/sp/HeRCCVYZ15}. Moreover, vendors may call misused APIs multiple times in firmware or reuse TPCs with misused APIs in different firmware, leading to aggravated TPC usage violation problems. 
Therefore, it is essential to detect TPC usage violations in IoT firmware.

\vspace{-0.3cm}
\subsection{Challenges}

To detect TPC usage violations in IoT firmware automatically, we have the following key challenges. 

\noindent\textbf{API specification inference from unstructured TPC document.}  
API specifications describe the requirements for using the API, which is critical for determining if the API usage is correct or wrong. Nevertheless, API specifications are typically implicitly encoded or explicitly presented in massive TPC documents, which are usually loosely organized and unstructured, hindering the API specification inference. Therefore, the first challenge is to effectively and precisely obtain the specifications of all the APIs in TPCs, which is an essential step to enable the usage violation detection.
A series of works have leveraged NLP techniques to extract API specifications from TPC documents~\cite{DBLP:conf/icsm/PanditaTWT16,DBLP:conf/icse/PanditaXZXOP12,lv2020rtfm}. Nevertheless, most of them only perform well on well-formatted TPC documents and are hard to handle unusual or ambiguous API specifications.
Besides, several works are proposed to infer API specifications through a lot of usage examples~\cite{DBLP:conf/uss/YunMSJKN16,DBLP:conf/kbse/KangRJ16}. However, the usage examples may be incorrect, not to mention it is hard, if not impossible, to cover all the usage cases of an API, which will result in inaccurate API specifications and introduce many false positives, causing low precision.

\noindent\textbf{Programming expression generation.} API specifications extracted from TPC documents are typically natural language, which cannot be directly applied to the usage violation detection. 
For instance, the API specification \textit{``the X509 object must be explicitly freed using X509\_free"} could be easily understood by the human but cannot be handled by the program without extra effort. Therefore, it is vital to translate the natural language-based API specifications into programming expressions, which will be used as input to the usage violation detection system. Nevertheless, it is not trivial to design a practical method to generate programming expressions from API specifications automatically.


\noindent\textbf{Usage violation detection.} In practice, IoT firmware images are mostly distributed as the closed-source binaries. Therefore, the third challenge is to perform the usage violation detection at the binary-level.
Previous works focus on the source-level API misuses. Therefore, they can rely on existing static analysis tools, e.g., \textit{CodeQL}~\cite{codeql}, or straightforward checkers to conduct analysis. Nevertheless, detecting the usage violation at the binary-level is much more complicated than that at the source-level, which will further cause challenges. First, the binary loses much information when compiling from the source code, e.g., missing symbols in the stripped binary, which is essential for the usage violation detection. Second, it is challenging to obtain the exact operations related to API usage at the binary-level due to the complex logical relationships between assembly instructions. Besides, the different architectures used in the firmware, such as x86, ARM, and MIPS, lead to different kinds of assembly instructions, increasing the difficulty of analyzing firmware. Currently, there is no such analysis tool like \textit{CodeQL} for the binary-level detection.



\vspace{-0.2cm}
\subsection{Methodology} In this paper, we aim to address these challenges to detect TPC usage violations at scale. 
To this end, we design and implement \system, the first automated and practical  framework to conduct the TPC usage violation detection on binary IoT firmware. Our design philosophy is as follows.

\textbf{First}, to solve the challenge of inferring  API specifications, we start from designing a sentiment-based model to filter out irrelevant API descriptions from TPC documents by leveraging a coreference resolution model and a customized BiLSTM model with the multi-head self-attention mechanism.  Next, we propose a Machine Reading Comprehension (MRC)-driven approach to extract API specifications from the processed TPC document by customizing an MRC system with our well-designed query questions and the manually constructed dataset.
\textbf{Second}, to address the problem of generating programming expressions, we design an NLP-guided
approach by constructing the dependency tree for each API specification according to the part of speech (POS) and then map the phrases in nodes into programming expressions based on semantic patterns, which are collected from previous works and enriched by using synonym replacement and changing the voice of patterns.
\textbf{Third}, we adopt \textit{FirmSec}~\cite{zhao2022large} for firmware processing, incorporating various enhancements. \textit{FirmSec} supports extracting multiple kinds of firmware and recognizing the TPCs in firmware by leveraging syntactical features and control flow graph (CFG) features. \textbf{Finally}, to solve the challenge of detecting usage violations in IoT firmware, we design a rule-driven analysis engine with customized violation-targeted checkers. The main idea behind our design is to encode the binary code into Datalog facts~\cite{dbiteboul1994foundations} and extract the corresponding implicit and explicit contextual logic relationships.  By parsing programming expressions into our analysis engine,  we successfully detect TPC usage violations in IoT firmware with high precision and recall.




\vspace{-0.3cm}

\subsection{Contributions} 
We summarize our main contributions as follows.

$\bullet$ We propose \system, the first automated and practical system to detect  TPC usage violations in binary IoT firmware, which fills the gap in mapping the high-level specifications from TPC documents to the binary-level violation analysis. 
\system achieves over 70\%  precision and recall in detecting TPC usage violations on ground-truth datasets, including 146 usage violations in three architectures. Though \system targets binary-level, it has an excellent performance improvement even compared to state-of-the-art source-level works: \textit{Advance}~\cite{lv2020rtfm}, \textit{APISAN}~\cite{DBLP:conf/uss/YunMSJKN16}, and \textit{APEx}~\cite{DBLP:conf/kbse/KangRJ16}. To facilitate future IoT security research, we will open-source \system at  \url{https://github.com/BBge/IoT-CVE}.

$\bullet$ To the best of our knowledge, we are the first to work on the TPC usage violation problem in IoT firmware and conduct the first large-scale analysis on this problem. Based on \system, we detect 27,621 usage violations caused by four TPCs in 4,545 firmware images. According to the results, we uncover the widespread TPC usage violations in IoT firmware. 
Our further case studies, the Denial-of-Service attack and the Man-In-The-Middle attack on several firmware images, demonstrate the  potential serious risk of TPC usage violations.
Up to now, 206 usage violations have been confirmed by vendors as vulnerabilities and seven of them have been assigned CVE IDs with high severity.


\vspace{-0.3cm}
\section{Background}
\vspace{-0.1cm}
\subsection{NLP Technique}
The NLP technique is an essential cornerstone in our system. Our NLP-based API specification extraction framework and NLP-guided programming expression generation approach involve multiple NLP concepts. We give a brief introduction to these concepts as follows.

\textbf{Coreference resolution} is a task that aims to automatically identify all mentions that refer to the same entity in a given text. Entities may be various kinds of names (e.g., API names), dates, locations, etc. Mentions are pronouns, named entities, and noun phrases. Currently, coreference resolution has lots of applications, including machine translation, full-text understanding, and so on. In this paper, coreference resolution is an essential step for extracting API specifications from documents since a TPC document usually has many entities, e.g., the return value and the arguments of an API.

\textbf{Multi-head self-attention} is derived from the self-attention mechanism which applies self-attention multiple times in parallel. The self-attention mechanism is a variant of the attention mechanism that relies less on external information and is better at capturing the inter-word dependencies of the sentence itself, such as common phrases, and entities referred to by pronouns. In this paper, the multi-head self-attention mechanism serves as an essential role in our customized sentiment-based model in extracting relevant API descriptions from TPC documents.

\textbf{Machine Reading Comprehension (MRC)} is a hot topic in NLP and has many real-world applications, e.g., information retrieval. An MRC system is designed to enable computers to understand the semantics of text and respond to natural language questions with precise answers by retrieving and reasoning through relevant knowledge, which is usually implicitly encoded or explicitly presented in the text. In this paper, we design an MRC-driven extraction method to extract precise API specifications from relevant API descriptions with our well-designed query questions and the manually annotated dataset.

\subsection{Datalog}
Datalog~\cite{dbiteboul1994foundations} is a declarative programming language that has been used in many tasks, including dataflow analysis~\cite{smaragdakis2010using}, binary disassembly~\cite{flores2020datalog}, and program trace encoding~\cite{li2021arbitrar}. Datalog utilizes the first-order predicate logic to represent the computation of logical relations. A Datalog program has two parts, an intensional database and an extensional database. The intensional database is composed of a set of Datalog rules while the extensional database is defined by Datalog facts. A Datalog rule is in the form of Horn clauses: $ Y :- X_1, X_2,..., X_n$, where $Y$ is the head of the rule, and $X_1, X_2,..., X_n$ are literals in the body of the rule. The above rule can be translated into natural language as ``If $X_1$ and $X_2$ and $...$ and $X_n$ are true, then $Y$ is true''. The rule head can be one or more predicates and the rule body can be predicates, negative predicates, or constraints. A Datalog fact is a special type of datalog rule, without any rule body, but only having a rule head.

Currently, Datalog has many variants, including \textit{Souffl{\'e}}~\cite{jordan2016souffle}, \textit{CodeQL}, and so on. \textit{Souffl{\'e}} is a new  logic programming language developed from Datalog, which overcomes several limitations in classical Datalog and has been adopted by many previous works. For instance, Flores-Montoya et al.~\cite{flores2020datalog} designed a binary disassembler based on \textit{Souffl{\'e}}. In this paper, we generate Datalog facts and rules for IoT firmware in \textit{Souffl{\'e}} syntax.

\vspace{-0.5cm}

\subsection{Motivation}
We discuss the motivation of this paper from three perspectives as follows.

\noindent\textbf{IoT vendor perspective.} In addition to using open-source TPCs, IoT vendors may also integrate closed-source TPCs in their IoT firmware. However, even closed-source TPCs can have serious security vulnerabilities resulting from API misuses. For example, the closed-source Qualcomm QCMAP suite, which is widely used in routers and IoT gateways, has been found to have a vulnerability due to the lack of checking the return values of functions, leading to the Denial-of-Service attack~\cite{qcmapvul}. Therefore, it is essential for IoT vendors to thoroughly evaluate and test closed-source TPCs for API misuses to ensure the security of their devices.

\noindent\textbf{IoT security company perspective.} Many IoT vendors turn to specialized IoT security companies to perform security assessments of their products. To maintain the privacy and security of their data, IoT vendors typically only provide closed-source firmware to these security firms. To conduct a comprehensive security evaluation of firmware, it is crucial for these security companies to have the ability to detect TPC usage violations within the closed-source firmware.

\noindent\textbf{Consumer perspective.} Due to the widespread use of IoT devices and the ever-emerging vulnerabilities, consumers (e.g., researchers and business companies) are becoming increasingly concerned about the security of IoT devices. TPC usage violations in IoT firmware can result in vulnerabilities that may lead to privacy breaches and significant losses for consumers.  As consumers usually only have access to the closed-source firmware of their products, it is important for those concerned about IoT product security to have a practical system that can detect TPC usage violations in IoT firmware.

\section{\system Design}
\label{section: Design}
\begin{figure*}
\centering
\includegraphics[width=\textwidth]{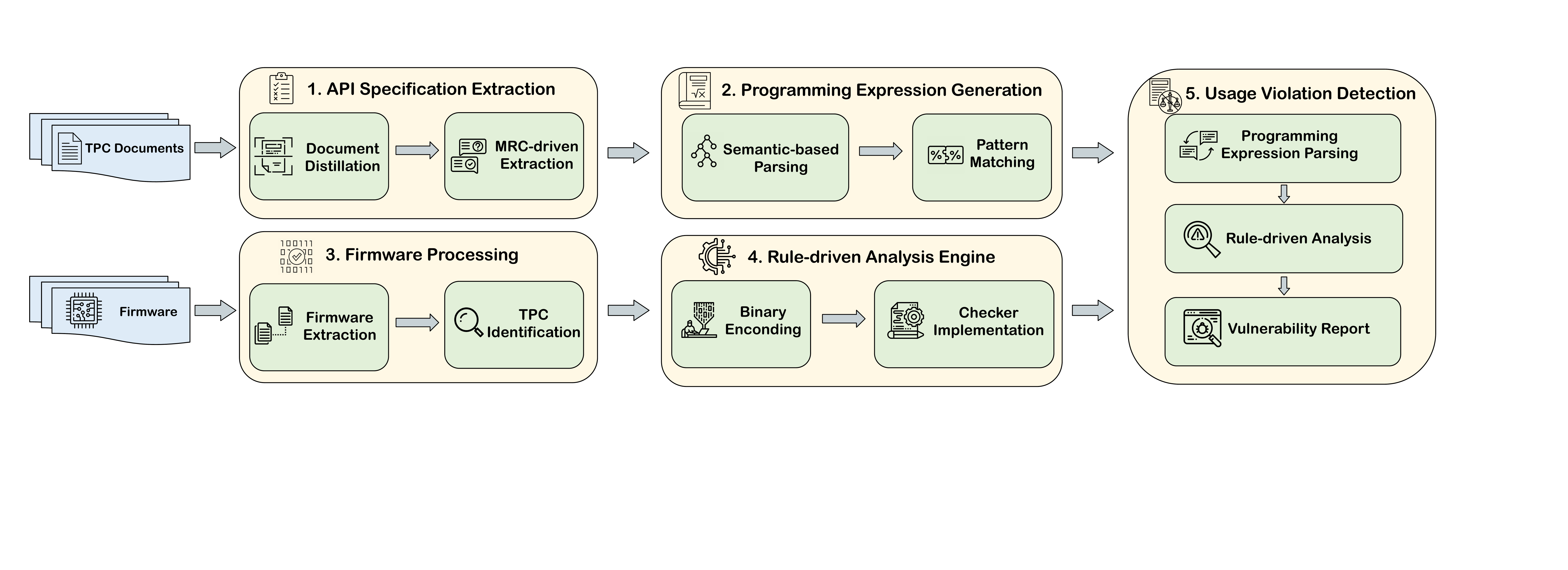}
\caption{Framework of \system.}
\label{Figure: system framework}
\end{figure*}

In this section, we present the  design details of \system.  At a high level, \system aims to automatically find TPC usage violations in IoT firmware. 
As shown in Figure~\ref{Figure: system framework}, \system mainly has five modules: API specification extraction, programming expression generation, firmware processing, rule-driven analysis engine, and usage violation detection.
The workflow is as follows.

 \textbf{First}, the API specification extraction module accepts TPC documents as input and extracts API specifications for TPCs based on a customized sentiment-based model and an MRC-driven system. 
\textbf{Then}, the programming expression generation module translates API specifications into programming expressions. \textbf{Next}, the firmware processing module  extracts objects from firmware and identifies the TPCs used in firmware. The rule-driven analysis engine module accepts the extracted objects as input, encodes them into Datalog facts, and designs four usage violation checkers.
\textbf{Finally}, the usage violation detection module parses programming expressions into the rule-driven analysis engine and then performs the usage violation check on IoT firmware.


\vspace{-0.2cm}
\subsection{TPC Usage Violation Category}
\label{Section: Usage Violation Categories}

Before we introduce the design of \system, we first introduce four representative categories of TPC usage violations, which are summarized by analyzing many TPC usage violations reported in previous works~\cite{li2021arbitrar,lv2020rtfm,DBLP:conf/compsac/GuWL0019,DBLP:conf/uss/YunMSJKN16} and/or in the real-world applications. 

\noindent\textbf{Deprecated API violation.}
A set of APIs will be deprecated or abandoned during the development of TPCs for various reasons, e.g., security issues, low performance, etc. Using the deprecated APIs may bring serious security problems. For instance, attackers can break the cryptographic protection mechanisms if the target program is still using the deprecated API \texttt{RAND\_pseudo\_bytes()} from OpenSSL~\cite{CVE-2015-8867}.




\noindent\textbf{Return value violation.} A great number of APIs have return values, which are used for indicating function execution results or status. If the return value of an API is unchecked or incorrectly checked, it may bring unpredictable security problems. For instance, it has been found that several functions for policy enforcement in Apache Accumulo do not properly check the return value will cause authorization bypass~\cite{CVE-2020-17533}.


\noindent\textbf{Argument violation.} When many APIs are invoked, the corresponding arguments should also be passed into APIs. These arguments often have strict constraints. For instance, the \texttt{errbuf} argument in \texttt{pcap\_open\_live()}, which is from libpcap, should be set into a zero-length string before calling the API. Failure to check the argument of an API can have serious consequences. For instance, the Zephyr Project, a scalable real-time operating system, did not validate the argument in some system calls in versions 2.1.0 and later versions prior to 2.2.0, causing the privilege escalation attack~\cite{CVE-2020-10058}.

\noindent\textbf{Causality violation.} Many APIs may have a strict causal relationship.  For instance, lock and unlock, fopen and fclose, as well as malloc and free are the three most common causal relationships. Besides, many APIs also have pre- and post-condition requirements, which we also regard as causal relationships. For instance, \texttt{sqlite3\_config()}, which is from SQLite, should be called before \texttt{sqlite3\_initialize()} or after \texttt{sqlite3\_shutdown()}.
Violating the required causal relationship may cause critical consequences, e.g., information leakage and system crash.



\vspace{-0.2cm}
\subsection{API Specification Extraction}
\label{section: API Specification Extraction}
The purpose of API specification extraction is to extract the API specifications from corresponding TPC documents, which will be used in the usage violation detection module.


\begin{figure}
\centering
\includegraphics[scale=0.4]{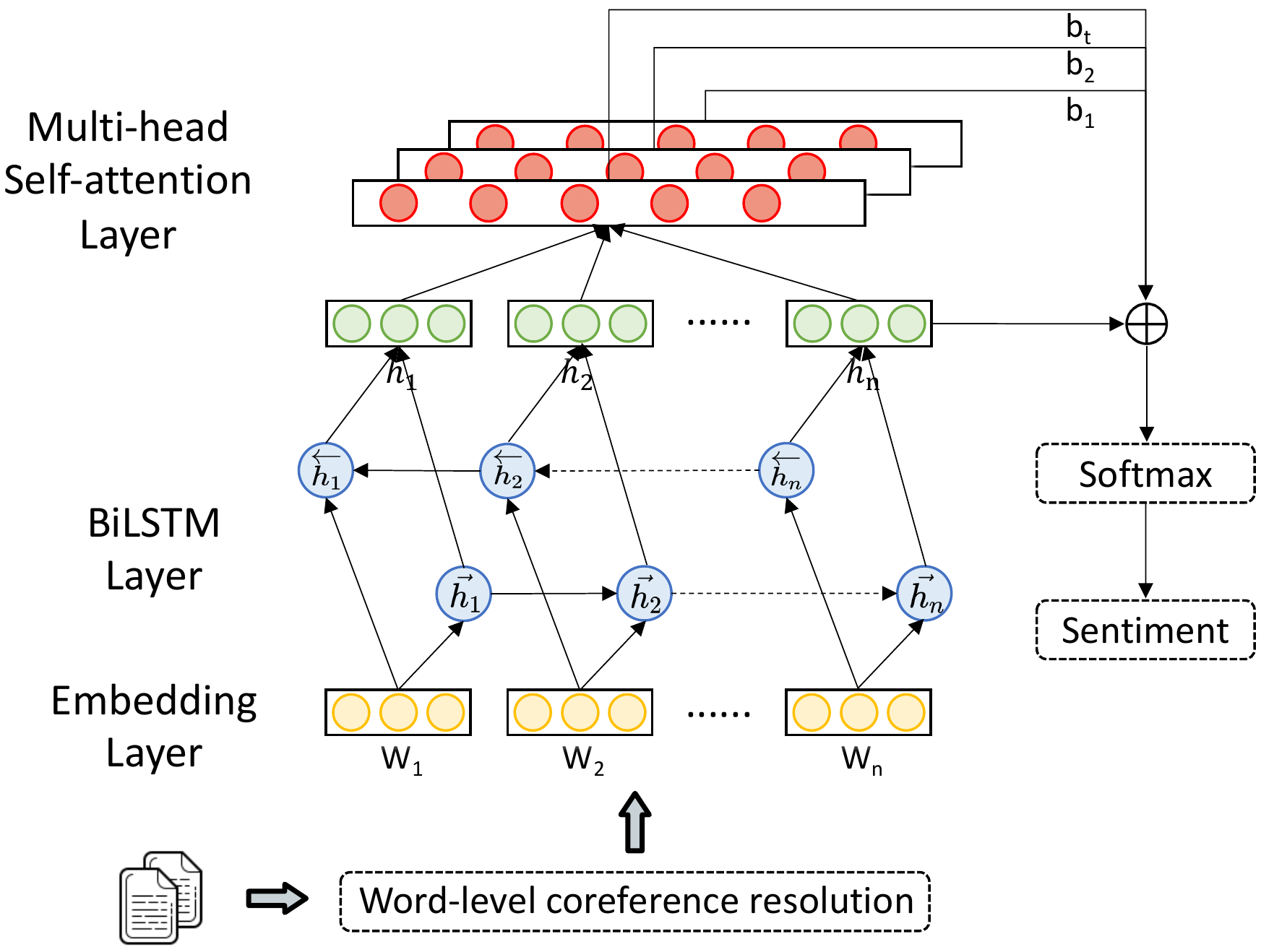}
\caption{Sentiment-based document distillation model.}
\label{Figure: Model Architecture}
\vspace{-0.65cm}
\end{figure}

\vspace{-1em}
\subsubsection{Document Distillation} TPC documents typically contain a wealth of sentences describing various aspects of the APIs. In this paper, we regard each sentence in the TPC document as an API description. Nevertheless, not all API descriptions are relevant to API specifications, which we regard as irrelevant API descriptions. Intuitively, these irrelevant API descriptions may affect the reliability of the final extracted API specifications.
 Therefore, to precisely extract API specifications, we should first distill TPC documents by filtering out irrelevant API descriptions and digging for relevant API descriptions. 
Unfortunately, TPC documents are always loosely organized and do not have consistent formats. It is difficult to recognize the relevant API descriptions automatically and precisely.

Previous research indicates that relevant API descriptions usually have a strong sentiment~\cite{lv2020rtfm}. For instance, a description in OpenSSL states that ``\textit{the initialization vector iv should be a random value}," which has a strong sentiment. Nevertheless, this observation does not apply to all scenarios when analyzing the TPC document and will introduce false positives. 
For example, the sentence ``\textit{additionally it indicates that the session ticket is in a renewal period and should be replaced}'' has a strong sentiment word ``\textit{should}'' but it is not a relevant API description. In this case, the pronoun ``\textit{it}'' actually represents the return value of the API, and the sentiment word ``\textit{should}'' refers to ``\textit{session ticket}'', which is not related to the specification of the API. 
The main reason behind this problem is that the above observation does not consider that the sentiment is target-dependent. There are two challenges to address this problem. First, we need to resolve the coreference in API descriptions. Second, we need to design a practical method to distinguish the sentiment of different targets and obtain the exact sentiment of the API.

To address the above challenges, we propose a novel sentiment-based document distillation model, leveraging the coreference resolution model, the bidirectional long-short term memory network (BiLSTM)~\cite{2015bilstm} with the multi-head self-attention mechanism~\cite{voita2019multihead}. Specifically, the coreference resolution model is designed to resolve the coreferences, e.g., pronouns, in TPC documents. BiLSTM performs well in learning the hierarchical information of sentences. The multi-head self-attention mechanism can capture strong sentiment words by leveraging the contextual information as well as the syntactic and semantic features. Since BiLSTM can enhance the semantic abstraction ability of the multi-head self-attention mechanism, we customize a sentiment analysis network by combining these two techniques, which can obtain the sentiment of each target and find the exact sentiment of the API.


Figure~\ref{Figure: Model Architecture} presents the architecture of our model. We first leverage WL-Coref~\cite{21coreference}, an off-the-shelf coreference resolution model, to resolve the coreferences in the TPC document. Next, we pass the processed TPC document into our BiLSTM model with the multi-head self-attention mechanism. The sentence will be converted to word vectors in the embedding layer. Then, the BiLSTM layer will analyze the sentence from both forward and backward directions and extract hidden vectors from word vectors. Hidden vectors will be passed into the multi-head self-attention layer. Finally, the output of the multi-head self-attention layer will be fed into a softmax function to obtain the final sentiment of a sentence. If the sentence has a strong sentiment, we regard it as a relevant API description.

In addition, three authors of this paper manually label relevant API descriptions to construct the dataset for training and evaluating our sentiment-based document distillation model. Though strong sentiment words (e.g., ``\textit{must}" and ``\textit{should}") are good criteria to label relevant API descriptions, we not only rely on them since they may lead to false positives (as discussed in previous paragraphs) and false negatives. An example of the false negative is the sentence ``\textit{call SSL\_get\_error() with the return value ret to find out the reason},'' which lacks strong sentiment words but is still considered a relevant API description. This sentence also has a strong sentiment since it is imperative and gives a forceful command. Therefore, we label the relevant API descriptions according to a set of criteria, including strong sentiment words, sentence patterns, negative structures, etc. The results annotated by each author will be cross-checked. We describe the dataset details in Section~\ref{Section: dataset}.









\begin{table}[]
\setlength{\abovecaptionskip}{0.02cm}
\caption{Question set.}
\setlength\tabcolsep{3pt}
\footnotesize
\label{Table: Question sets}
\resizebox{0.48\textwidth}{!}{
\begin{tabular}{c|c}
\toprule[1.5pt]
\bfseries{Category}                                                                & \bfseries{Question}                                                              \\ \midrule[1pt]
\multirow{2}{*}{\begin{tabular}[c]{@{}c@{}}Return\\ Value\end{tabular}} & What are return values supposed to be?                                \\
                                                                        & In which condition does the function have a return value?            \\ \midrule[1pt]
\multirow{4}{*}{Causality}                                              & What operation is required if the return value is $ReturnValue_{i}$?                       \\
                                                                        & What operation is required if $Condition_{i}$?                        \\
                                                                        & Which function should be called before the API?                           \\
                                                                        & Which function should be called after the API?                 \\ \midrule[1pt]
\multirow{4}{*}{Argument}                                               & What is the value of the N-th argument supposed to be before the API? \\ 
                                                                        & What is the value of the N-th argument supposed to be after the API?  \\ 
                                                                        & How to check the N-th argument before the API?                        \\ 
                                                                        & How to check the N-th argument after the API?                         \\ \bottomrule[1.5pt]
\end{tabular}
}
\end{table}

\subsubsection{MRC-driven Extraction}
In this step, we extract precise API specifications from relevant API descriptions, which will be further converted into programming expressions and used by the rule-driven analysis engine. 
Nevertheless, it is challenging to extract precise API specifications since it usually involves contextual understanding which cannot be properly handled by the keyword or template matching~\cite{pandita2012inferring, DBLP:conf/icsm/PanditaTWT16}.
For instance, a typical API description could be ``\textit{SQLITE\_OK be returned by sqlite3\_snapshot\_recover if successful, or an SQLite error code otherwise}''. For this case, we need to understand the context to infer that the latter part of the sentence is equal to  ``\textit{SQLite error code be returned by sqlite3\_snapshot\_recover if failed}''. 
To address the problem, we propose a novel MRC-driven extraction method by using the MRC system to extract precise API specifications. There are two main challenges to apply the MRC system to our task. First, there are no relevant examples to guide the design of query questions about TPC usage violations, which are an important part of the MRC system. Second, there is no publicly accessible dataset that can be used to train our MRC system. A well-labeled dataset is important for the performance of an MRC system.
We elaborate our solutions to these challenges as follows.

\noindent\textbf{Query question design.} To solve the challenge of designing query questions, we analyze the possible causes of each kind of TPC usage violation and design the corresponding questions. For instance, the possible causes of the causality violation are lacking a call to the required function before or after the API, and missing or incorrect operations under certain conditions or return values. Therefore, we design query questions for the causality violation based on the above possible causes. We do not design questions for the deprecated API violation since we can obtain the deprecated API list for the TPC from its official website.

Table~\ref{Table: Question sets} presents the query questions designed for different kinds of usage violations mentioned in Section~\ref{Section: Usage Violation Categories}. These questions are well-suited for our MRC system for three key reasons. First, these questions are summarized through a thorough analysis  of hundreds of real-world TPC usage violations and the examination of numerous TPC documents. They are all highly relevant to the possible causes of various usage violations and cover the vast majority of cases. Second, these questions contain both extractive and no answer questions, where the answer to each question is either a continuous subsequence extracted from API descriptions or no answer. Developing the ability to answer such questions, especially those with no answer, can enhance the system's ability to understand and interpret the context, further improving its generalization ability when presented with new TPC documents. Finally, all questions are clearly phrased and straightforward, rendering them suitable for further dataset annotation.

Additionally, the answers to two questions related to the return value will become part of the two questions related to the causality ($Condition_{i}$ and $ReturnValue_{i}$).  $Condition_{i}$ indicates the prerequisites of an API having the return value, which is often closely linked to $ReturnValue_{i}$. Typically, TPC documents specify the required operations that correspond to specific return values, and these operations are presented together with the respective return values.  Nevertheless,  there are cases where the required operations are not associated with the return value, but instead with the condition. For instance, libpcap documentation states that ``\textit{pcap\_activate returns 0 on success without warnings, a non-zero positive value on success with warnings,  and a negative value on error. If pcap\_activate fails, the pcap\_t $*$ is not closed and freed; it should be closed using pcap\_close}". In this scenario, the operation (\textit{close pcap\_t $*$ using pcap\_close}) is associated with the condition (\textit{pcap\_activate fails}) rather than the return value (\textit{a negative value}). Therefore, in light of the above situation, we design specific query questions to obtain the exact operations under certain conditions. 

\begin{figure}[H]
\centering
\includegraphics[scale=0.38]{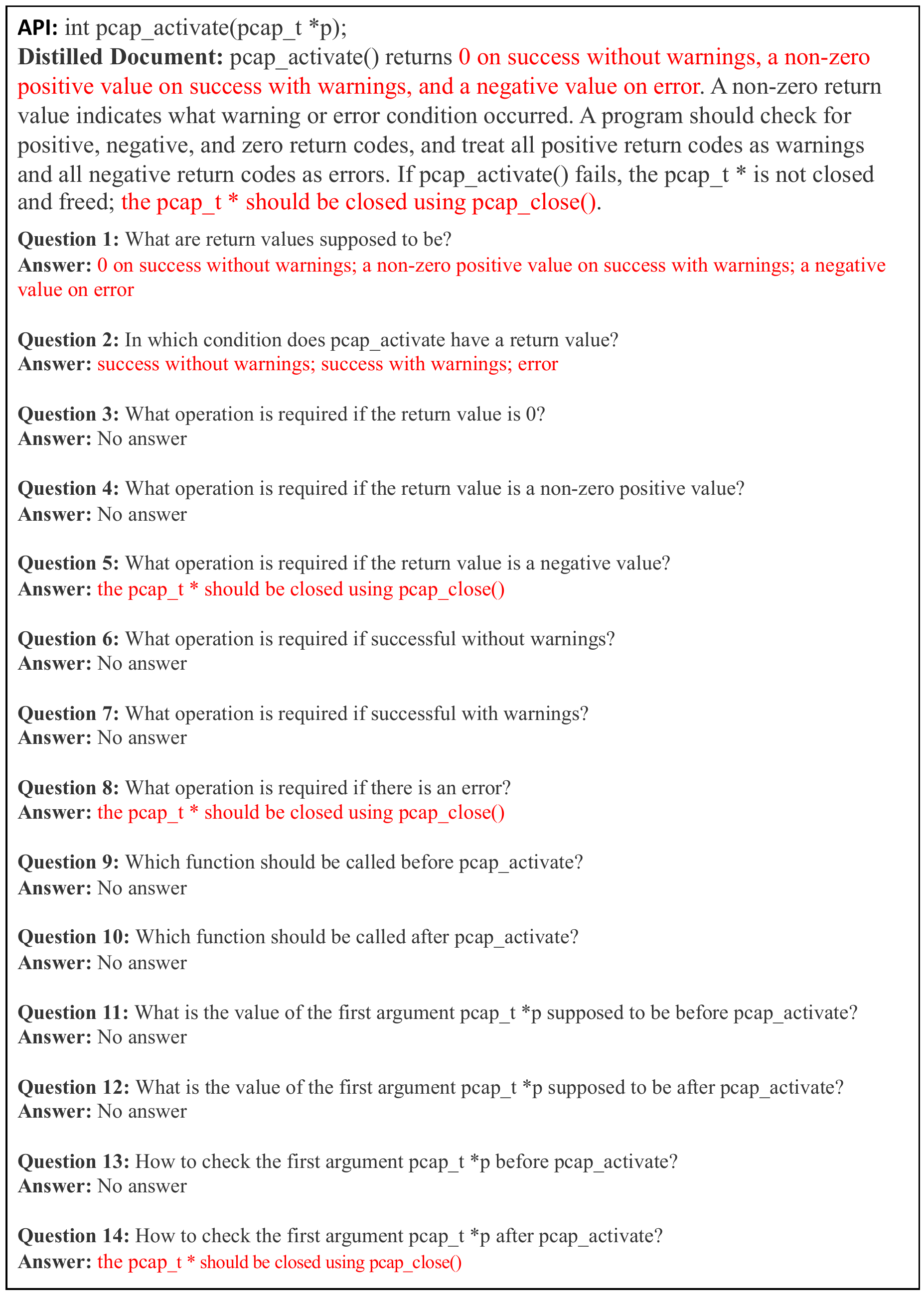}
\caption{An example of a distilled document with associated query questions and annotations.}
\label{Figure: answer annotate}
\end{figure}

\noindent\textbf{Dataset construction.} To address the problem of lacking the dataset, we manually review 15,000 API descriptions from TPC documents and annotate the answer to each designed question, which is a one-time task.
Specifically, most answers are annotated with the continuous subsequence since we need the auxiliary information to generate the programming expression later. For questions that do not have an answer, such as those where the document does not necessitate further operations under certain return values, they are marked as unanswerable.  Additionally, the annotated condition will be slightly rephrased to fill in the stem of the corresponding question ($Condition_{i}$). The rephrasing is performed automatically by Grammarly, a well-known grammar checker.
Take the descriptions of libpcap in the last paragraph for example. We first annotate the ``\textit{success with/without warnings}" and ``\textit{error}" as conditions. Next, we directly put the condition into the stem of the question and rephrase it by using Grammarly. The final question will be formed as ``\textit{What operation is required if successful with/without warnings/there is an error?}'' The results annotated by each author will be cross-checked. We describe the dataset details in Section~\ref{Section: dataset}. Moreover, we provide an example of a distilled document with associated query questions and annotations, as shown in Figure~\ref{Figure: answer annotate}.

\noindent\textbf{MRC system implementation.} Currently, there are various off-the-shelf MRC systems.
By comparing different MRC systems, we finally choose to use \textit{RoBERTa}~\cite{liu2019roberta} in our system.
\textit{RoBERTa} is optimized from BERT~\cite{DBLP:conf/naacl/DevlinCLT19} and has been proved efficient in many MRC tasks~\cite{DBLP:conf/acl/RajpurkarJL18}. 
We fine tune the MRC system on our ground-truth dataset to obtain a great performance, which achieves an 88.23\% F1 score.

Additionally, it is important to note that we will check for overlapping answers to different questions in order to identify any shared preconditions and avoid false positives. For instance, as illustrated in Figure~\ref{Figure: answer annotate}, Questions 5, 8, and 14 share the same answer and have a common precondition, which is ``\textit{pcap\_activate() returns a negative value on error}." If we fail to identify this precondition for Question 14, the system may erroneously check \texttt{pcap\_t *} when \texttt{pcap\_activate()} succeeds, resulting in a false positive.

\vspace{-0.3cm}
\subsection{Programming Expression Generation} 
Since API specifications are in natural language, they cannot be directly understood by our analysis engine without some form of processing or translation. Therefore, it is necessary to create a structured, machine-readable representation of each API specification, which is nevertheless very challenging.  The main problem here is that API specifications do not have consistent formats and thus cannot be simply mapped by keywords or templates. 

\begin{figure}
\centering
\includegraphics[scale=0.4]{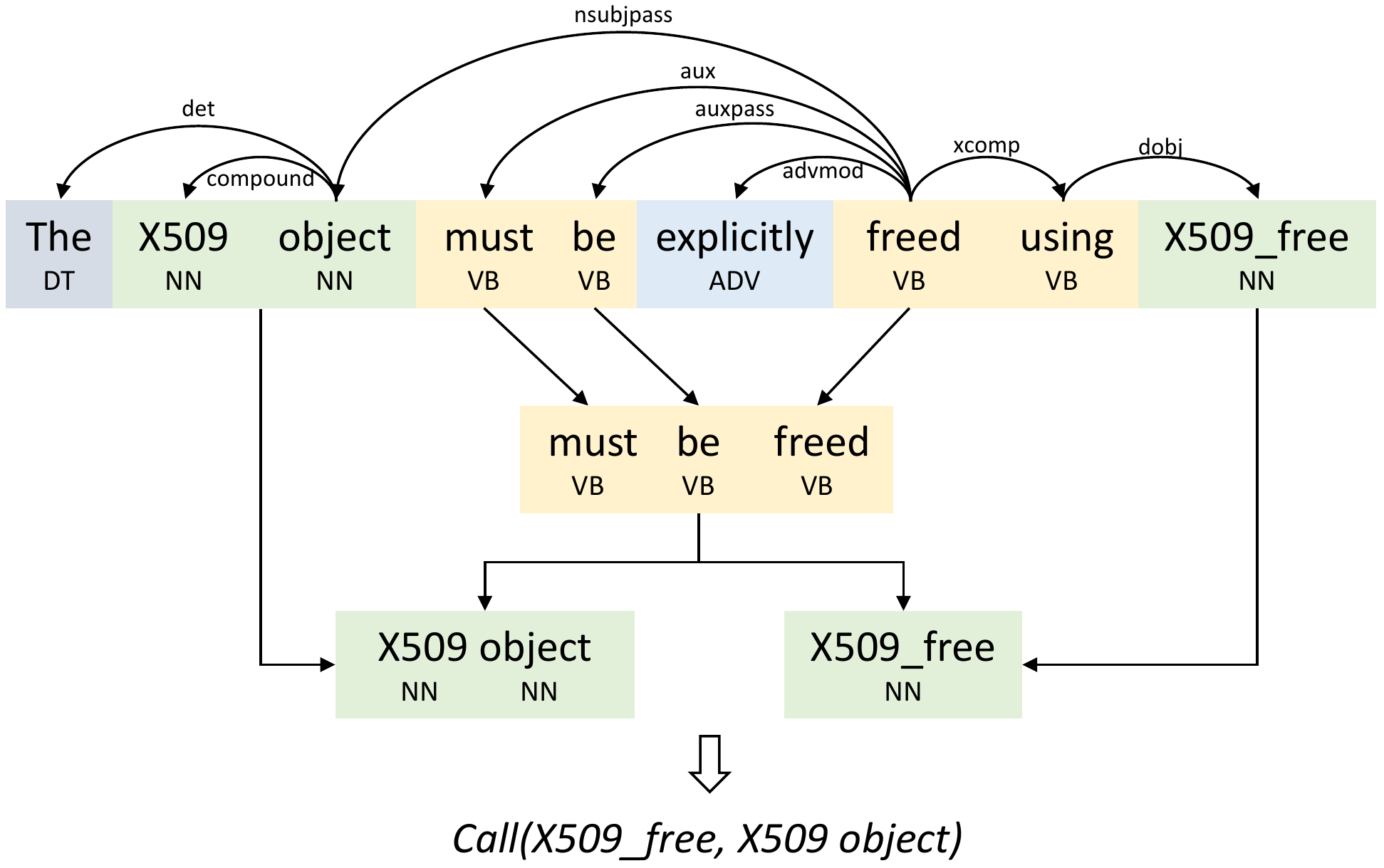}
\caption{An example of programming expression generation.}
\label{Figure: Programming Expression Generation}
\end{figure}

To address the problem, we design a novel NLP-guided approach by breaking the API specification into small phrases according to their POS, constructing the corresponding dependency tree, and then mapping phrases into programming expressions. These programming expressions are structured, machine-readable representations that follow a specific format, which we define as \textit{Operation(argument1, argument2, ...)}.
More specifically, \textbf{first}, we leverage the POS tagging~\cite{coreNLP2014stanford} to annotate each word in an API specification and identify the relation between the words. \textbf{Second}, we create a dependency tree by combining words with a close relationship. The root node describes the relation between the two leaf nodes. Some words, e.g., determiners and adjectives, will be removed from the dependency tree.
\textbf{Third}, we maintain a list that includes patterns to map common phrases into programming expressions, e.g., map ``greater than or equal to'' into ``$>=$''. The patterns consist of three parts. The first part is the patterns collected from previous works~\cite{blasi2018jdoctor,lv2020rtfm}. The second part is generated from the first part by using synonym replacement. We leverage WordNet~\cite{wordnet}, a large lexical database of English, to find the synonym of a pattern. For instance, ``\textit{must be freed}'' can be replaced as ``\textit{must be released}''. The third part is generated from the first and second parts by changing the voice of the patterns, e.g., from the passive to the active voice and vice versa, which will not change the meaning of patterns. \textbf{Finally}, we traverse the dependency tree in post-order, and map phrases in root nodes into programming expressions according to the patterns. In addition, to provide a clearer understanding of the process, we use the specification ``\textit{the X509 object must be explicitly freed using X509\_free}'' as an example, as shown in Figure~\ref{Figure: Programming Expression Generation}. Initially, we annotate the POS for each word in the specification and identify their relationships. Next, we construct a dependency tree that represents the structure of the specification. The root node ``\textit{must be freed}'' indicates the relationship between the two leaf nodes ``\textit{X509 object}'' and ``\textit{X509\_free}.'' Finally, the phrase ``\textit{must be freed}'' is matched by our patterns and translated into the predefined ``\textit{CALL}'' operation type. The two leaf nodes are used as arguments for the ``\textit{CALL}'' operation.


\vspace{-0.2cm}
\subsection{Firmware Processing}
The goal of firmware processing is to extract the contained objects and identify the target TPC in firmware.
We mainly process firmware by two steps: firmware extraction and TPC identification.
\vspace{-0.2cm}
\subsubsection{Firmware Extraction} The first step is to unpack firmware and extract the contained objects, including the linked libraries, filesystems, and so on. These objects are vital for later TPC identification and binary encoding. To achieve this goal, we mainly leverage the extraction module of \textit{FirmSec}~\cite{zhao2022large}, which is implemented based on \textit{binwalk}~\cite{binwalk} and provides extra support for three filesystems: SquashFS, JFFS2, and YAFFS. During our analysis, we find the original extraction module cannot deal with the UBI filesystem. To address this issue, we make an improvement to the extraction module by implementing a plugin that is specifically designed to extract objects from the UBI filesystem.

\vspace{-0.2cm}
\subsubsection{TPC Identification} 
The second step is to identify the target TPC in firmware and recover the unknown TPC-related APIs. Locating the TPC and its APIs in firmware is a basic requirement for further usage violation analysis. Generally, source-level API misuse detectors can easily find the target APIs by searching the source code. Nevertheless, it is challenging to locate the target APIs at the binary-level since firmware may be stripped and the symbols are removed. 
The main problem here is to identify the TPC and its APIs in firmware with limited information. 
To solve this problem, we mainly leverage \textit{FirmSec}, a state-of-the-art tool for TPC identification, with two enhancements. \textit{FirmSec} can recognize the TPCs used in firmware at TPC-level and version-level with high precision by using syntactical features (e.g., function names) and control-flow graph (CFG) features from TPCs and firmware. 
Specifically, \textit{FirmSec} first leverages the edit distance and ratio-based matching to determine the similarity of unstripped syntactical features. Next, \textit{FirmSec} uses a customized neural network to compare the CFG features. Finally, \textit{FirmSec} combines the results of the syntactical feature matching  and CFG feature matching to identify TPCs in IoT firmware.

Our enhancements are as follows. \textbf{First}, except for the original syntactical features (e.g., string literals and function names) used by \textit{FirmSec}, we adopt two extra syntactical features. \textit{B2SFinder}~\cite{feng2022finder} indicates that complex branch sequences (e.g., switch/case and if/else structures) are hardly changed during the compilation. The constant conditions used in switch/case and if/else structures will not be removed in the stripped firmware while string literals and function names may be removed. Therefore, we extract the constant conditions in switch/case and if/else structures as new syntactical features. \textbf{Second}, we simplify the CFG feature matching process by eliminating redundant CFG feature matching computations. 
We filter out some obvious irrelevant functions, such as having completely different function names and no corresponding special strings, in advance based on the syntactical features mentioned above and only perform the CFG feature matching on the remaining functions.

Additionally, though syntactical features are important to \textit{FirmSec}, it can still perform TPC identification when all syntactical features are stripped. Specifically, \textit{FirmSec} achieves a precision of around 90\% and a recall of 80\% on TPC identification by solely relying on CFG features~\cite{zhao2022large}. Therefore, the existence of CFG features is the minimum necessity to enable our TPC identification. Moreover, \textit{FirmSec} can recover the stripped function names according to CFG feature matching results. These recovered function names will be further used to locate APIs in IoT firmware.

\vspace{-0.3cm}
\subsection{Rule-driven Analysis Engine}
Though we have extracted API specifications from TPC documents and generated the corresponding programming expressions, we still lack a practical method to leverage them for usage violation detection. Existing API misuse detectors do not consider to use API specifications too much. 
This is because they mainly perform the detection at the source-level, and thus it is possible and much easier for them to rely on existing tools for the API misuse detection. 
For instance, \textit{Advance} uses a sophisticated system \textit{CodeQL}~\cite{codeql} to conduct API misuse detection.  Nevertheless, detecting the usage violation at the binary-level is much more complicated than that at the source-level. It is challenging to obtain the exact operations related to API usage at the binary-level due to the
complex logical relationships between assembly instructions. Currently, there is no solution that could properly handle this problem. Therefore, to address the challenge, we introduce Datalog to our task, which is powerful in logical inference. Nevertheless, there are also two challenges to apply Datalog to our task. First, it is challenging to extract the information in firmware about API usage and encode it into facts. Currently, there is no uniform standard for encoding firmware into facts. Second, it is difficult to design Datalog rules to perform the usage violation check on the generated facts since the logical inference for different usage violations varies widely. To solve the above challenges, we implement a rule-driven analysis engine with the following designs.

\vspace{-0.3cm}
\subsubsection{Binary Encoding}  To address the challenge of encoding the binary into Datalog facts, we develop a binary-to-Datalog encoding tool that takes advantage of  \textit{Ddisasm}~\cite{flores2020datalog},  along with our customized facts that mainly focus on the
register usage of an API.  The tool accepts the extracted binaries, with the identified TPC-related information, from IoT firmware as input, and encodes them into Datalog facts.

\begin{figure}
\centering
\includegraphics[scale=0.4]{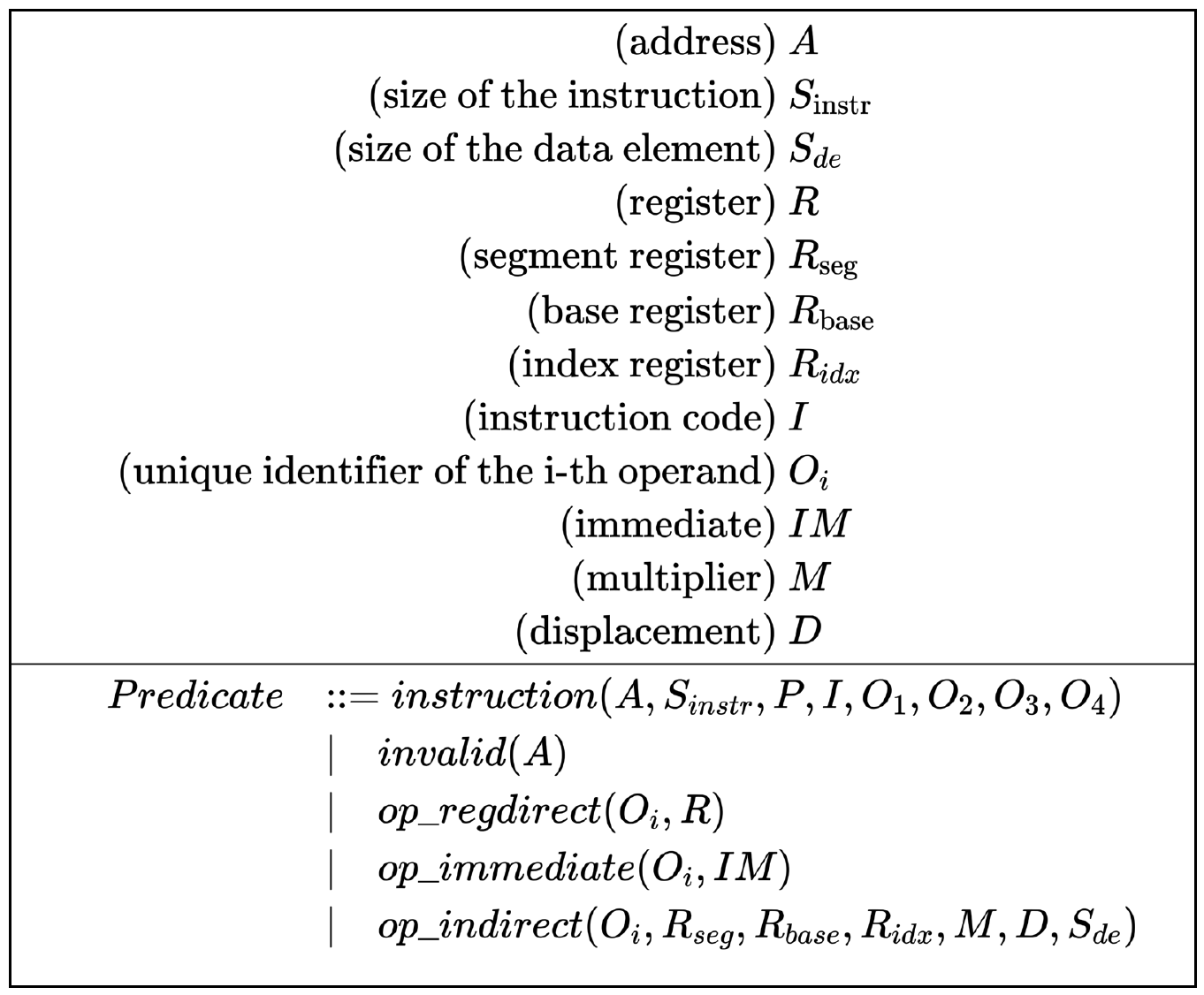}
\caption{Initial datalog facts used by \textit{Ddisasm}.}
\label{Figure: Datalog facts}
\end{figure}

\begin{figure}
\centering
\includegraphics[scale=0.4]{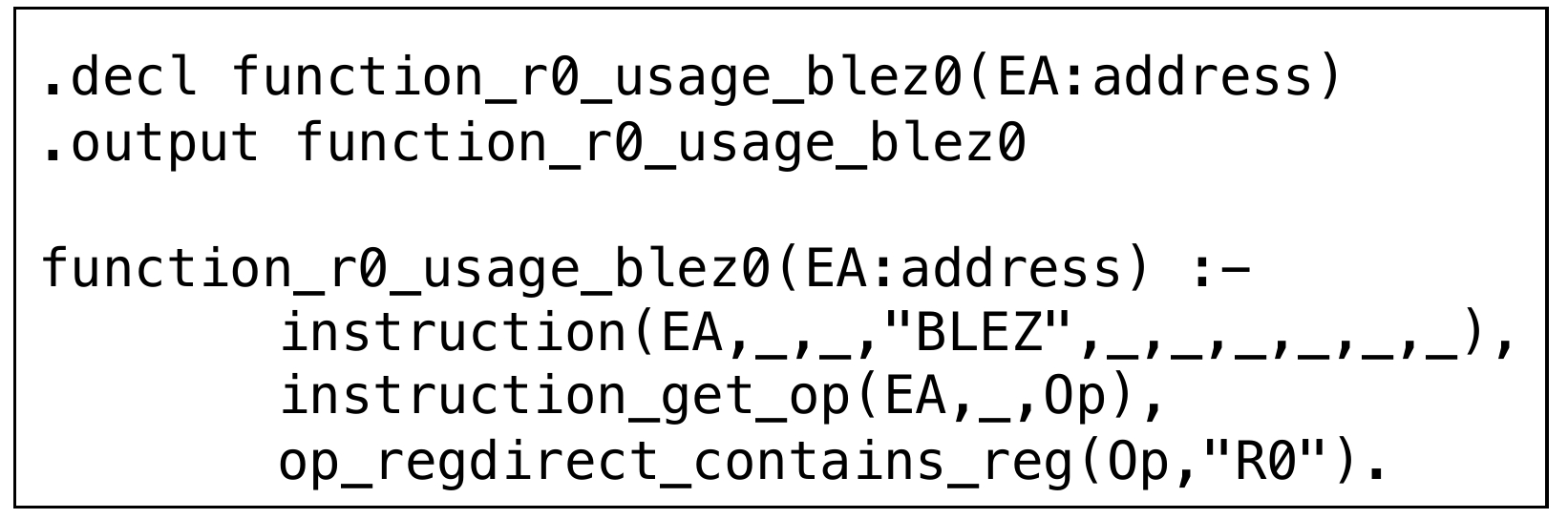}
\caption{An example of a rule that checks the return value.}
\label{Figure: factgeneration}
\vspace{-2em}
\end{figure}

\textit{Ddisasm}, a binary analysis and rewriting tool, has implemented  a method to encode the binary into Datalog facts.
  Figure~\ref{Figure: Datalog facts} presents the standard used by \textit{Ddisasm} of encoding each instruction in binary into a set of initial Datalog facts. Based on initial Datalog facts, \textit{Ddisasm} further generates more than 300 kinds of facts through various built-in rules. These facts represent various information in the binary. Nevertheless, we find only a few facts related to the API usage, hindering the TPC usage violation analysis. 
To solve this problem, we customize rules to generate facts that correspond to the usage of APIs, which are related to return values, arguments, and adjacent calling functions. Specifically, \textbf{first}, we generate facts describing the register that holds the return value or the argument of an API. We declare relations related to the register usage and then customize rules for each relation based on existing facts. The output of the relation is the generated fact.
For instance, we first declare a relation \texttt{function\_r0\_usage} to describe the R0 register usage, which holds the return value or the argument of a function in ARM32.  Next, we implement the rules based on the existing facts, e.g., \texttt{instruction\_get\_op} and \texttt{op\_regdirect\_contains\_reg} to obtain the corresponding facts of \texttt{function\_r0\_usage}. As shown in Figure~\ref{Figure: factgeneration}, we present an example of a rule that verifies whether the return value is checked using the condition $<=0$. The rule is based on three existing facts and the output of the rule is the fact \texttt{function\_r0\_usage\_blez0}. This rule can be interpreted as first finding the instructions that have the instruction code BLEZ and then further identifying those instructions that include the R0 register.
\textbf{Second}, we generate facts describing the adjacent function calls before or after an API. The function calls can be identified by focusing on certain instructions, e.g., branch instructions. Therefore, similar to the first step, we define rules to extract the function names in certain instructions. \textbf{Finally}, we generate facts describing the operations under certain return values, including function calls, value assignments, etc. In this step, we define rules to extract the function names in certain instructions and the register usage after the return value is checked.
The facts generated by the second and third steps will be further used to identify the causality violation of an API.

\vspace{-0.3cm}
\subsubsection{Checker Implementation}
To address the challenge of performing the usage violation check on the generated
facts, we design four checkers based on the features of different TPC usage violations. 

\noindent\textbf{Deprecated API violation checker.} We maintain a list of deprecated APIs for each TPC. Therefore, comparing the function names extracted from firmware with the deprecated APIs in the list is straightforward. Nevertheless, only comparing the function name may result in false positives. 
For instance, some firmware images may contain a self-implemented function with the same name as a deprecated API. Also, the different TPCs used in firmware may have functions with the same name. To solve this problem, we additionally check three features of the function when the function name matches the list, including the number of arguments, the argument types, and the possible function calling sequence. 




\noindent\textbf{Return value violation checker.} In this checker, we mainly focus on the operation of the registers that hold the return value of the function, e.g., the R0 register in ARM32. The checker accepts constraints on API return values from programming expressions as input. More specifically, we focus on two kinds of return value violations: lack of checking the return value and incorrect check of the return value.  In the former case, we check whether the specific register involves in compare instructions (e.g., CMP) or branch instructions (e.g., BEQ). In the latter case, we further check whether the compared value (e.g., 0) and the corresponding compare operation (e.g., BGE) are consistent with the constraints (e.g., greater than or equal to 0). It is important to mention that we do not simply restrict the scope of return value detection to the current caller function. Though the API is called in the current caller function, its return value may be checked in other functions. For instance, function A calls function B and checks its return value, while function B in turn calls the API and returns the return value of the API. Lack of consideration for this situation could result in false positives. We set the depth of the indirection return value check at most one function call.

\noindent\textbf{Argument violation checker.} Similar to the return value violation checker, we focus on the operation of the registers that hold the arguments of the function, e.g., the R0-R3 registers in ARM32. The checker accepts constraints on API arguments from programming expressions as input. Most constraints are related to the value of the arguments (e.g., the argument should be set to 0 before the call) and the handling of pointer arguments (e.g., the pointer argument should be freed after the call). Therefore, we first analyze the value of each argument and the operations of the pointer argument, before and after the call. Next, we check whether the values or operations are consistent with the constraints. In addition, similar to the indirection return value check, we also consider the argument that may be processed indirectly, e.g., free a pointer argument in a different function, to avoid false positives.

\noindent\textbf{Causality violation checker.} In this checker, we mainly focus on the functions that are called before or after the API, and the operations under certain return values. The checker accepts constraints on API causality from programming expressions as input. More specifically, we pay special attention to two kinds of causality violations: lack of calling the required function before or after the API (e.g., lock should be called before unlock), and 
incorrect or without operations under certain return values (e.g., call \texttt{SSL\_get\_error()} if \texttt{SSL\_do\_handshake()} returns 0).
In the former case, we define rules to obtain the five closest functions that are called before or after the API respectively. Our analysis indicates that the checker achieves optimal precision when the limit is set to five. We then check whether the required function is in the list of collected functions. 
In the latter case, we focus on the operations after the return value is checked, including function calls, value assignments, and so on. Similar to the return value violation checker, we do not restrict the scope of causality violation detection to the current caller function to avoid false positives.

The four checkers have multiple built-in rules that are plug-and-play and can be easily extended to new TPCs without requiring additional development efforts. This adaptability stems from the fact that the root causes for usage violations persist consistently across different TPCs, and all checkers are tailored in accordance with the specific root causes of usage violations. In addition, although the API specifications of various TPCs may differ, they will be standardized into a uniform format through programming expression generation.

\vspace{-0.2cm}
\subsection{Usage Violation Detection} The usage violation detection module aims to perform the Datalog checkers on the Datalog facts of binaries to recognize the potential TPC usage violations. In this module, we first implement a simple parser to pass programming expressions into Datalog checkers according to the operation type of programming expressions.
Next, we leverage \textit{Souffl{\'e}}, an off-the-shelf Datalog engine, to execute the Datalog checkers. Then, we set \textit{GHIDRA} as our front-end and present the visualizing results on it according to \textit{d3re}~\cite{sun2021d3re}.
According to the results, we finally generate a vulnerability report that indicates the recognized TPC usage violations in IoT firmware.





\vspace{-0.2cm}
\section{System Evaluation}
\label{Section: System Evaluation}

In this section, we first introduce the datasets used for evaluation. Next, we evaluate the performance of key components in \system and the overall performance of \system. We also compare \system with multiple state-of-the-arts.

\vspace{-0.2cm}
\subsection{Dataset}
\label{Section: dataset}
To enable our evaluation, we create the following six datasets.

\textbf{TPC documents dataset ($D_{TPC-DOC}$)} includes the latest documents from four TPCs: OpenSSL, SQLite, libpcap, and libxml2. More specifically, OpenSSL has 1,554 API calls, SQLite has 221 API calls, libpcap has 69 API calls, and libxml2 has 1,497 API calls.
We choose these four TPCs for concept validation for the following reasons. (1) They are widely used in IoT firmware according to the results presented in \cite{zhao2022large}. (2) They have the corresponding ground-truth usage violation dataset created by Lv et al.~\cite{lv2020rtfm}, and Gu et al.~\cite{DBLP:conf/compsac/GuWL0019}. 

\textbf{API description dataset ($D_{DESC}$)} is used for training and evaluating our sentiment-based document distillation model. It includes the randomly selected API descriptions from the aforementioned TPC documents. Specifically, $D_{DESC}$ contains $4,086$ API descriptions from OpenSSL and SQLite, including $927$ relevant API descriptions and $3,159$ irrelevant API descriptions. We split $D_{DESC}$ into three sets, training set ($D_{DESC_{train}}$), development set ($D_{DESC_{dev}}$), and testing set ($D_{DESC_{test}}$) respectively, according to the ratio of 6:2:2. In addition, to evaluate the generalization ability of our model, we construct an extra dataset ($D_{DESC_{new}}$) consisting of 733 API descriptions from libpcap and libxml2. $D_{DESC_{new}}$ includes $141$ relevant API descriptions and $592$ irrelevant API descriptions.

\textbf{MRC dataset ($D_{MRC}$)} is used for training and evaluating the MRC system of \system. It includes the manually labeled question-answer pairs for each API from OpenSSL and SQLite. More specifically, $D_{MRC}$ contains 19,536 question-answer pairs of 1,775 API calls. We split $D_{MRC}$ into three sets, training set ($D_{MRC_{train}}$), development set ($D_{MRC_{dev}}$), and testing set ($D_{MRC_{test}}$) respectively, according to the ratio of 6:2:2. Besides, we create an additional dataset ($D_{MRC_{new}}$), which includes 1,058 question-answer pairs of 63 API calls from libpcap and libxml2. The APIs in $D_{MRC_{new}}$ have the corresponding manually distilled documents in $D_{DESC_{new}}$. We use $D_{MRC_{new}}$ to evaluate the generalization ability of the MRC system and conduct the error propagation analysis.

\textbf{Programming expression dataset ($D_{PE}$)} contains the ground-truth programming expressions. We randomly select relevant API descriptions from the aforementioned four TPC documents and manually label the corresponding programming expressions. More specifically, $D_{PE}$ includes $600$ programming expressions.


\textbf{Real-world usage violation dataset ($D_{Real-UV}$)} includes the known usage violations, which are in popular C programs, that correspond to the aforementioned four TPCs. 
The purpose of $D_{Real-UV}$ is to evaluate the usage violation detection accuracy of \system. As shown in Table~\ref{Table: Ground-truth Usage Violation Dataset}, $D_{Real-UV}$ includes 77 known usage violations collected from two ground-truth datasets: 59 usage violations in 24 programs from \textit{Advance}~\cite{lv2020rtfm}, and $18$ usage violations in $2$ programs from \textit{APIMU4C}~\cite{DBLP:conf/compsac/GuWL0019} (the gray lines in Table~\ref{Table: Ground-truth Usage Violation Dataset}).

\textbf{Artificial usage violation  dataset ($D_{Artif-UV}$)} includes the manually created usage violations. The purpose of this dataset is to enlarge $D_{Real-UV}$ and provide a comprehensive evaluation of \system. We create $D_{Artif-UV}$ by manually inserting the incorrect code into the open-source IoT firmware, which includes $69$ manually created usage violations from $23$ firmware images, as shown in Table~\ref{Table: Ground-truth Usage Violation Dataset}.








\vspace{-0.4cm}
\subsection{Evaluation of \system}




\begin{table}[]
\caption{Ground-truth usage violation dataset.}
\label{Table: Ground-truth Usage Violation Dataset}
\centering
\footnotesize
\renewcommand\arraystretch{0.6}
\begin{tabular}{c|c|c|c} 
\toprule[1.5pt]

\multicolumn{1}{c|}{\bfseries{TPC}}                      & \bfseries{Program}                       & \bfseries{Version}  & \bfseries{\begin{tabular}[c]{@{}c@{}}\# Usage\\Violation\end{tabular}}   \\ 
\midrule[1pt]
\multicolumn{4}{c}{\boldmath $D_{Real-UV}$}                                                                                      \\ 
\midrule[1pt]
\multirow{14}{*}{OpenSSL} & \multirow{2}{*}{dovecot}      & 0eaf77d  & 1                                                         \\
                          &                               & 394391e  & 1                                                         \\ \cmidrule{2-4}
                          & mutt                          & 101e05d6 & 6                                                         \\ \cmidrule{2-4}
                          & \multirow{2}{*}{ntp}          & 2383333  & 1                                                         \\
                          &                               & c70fc4b  & 1                                                         \\ \cmidrule{2-4}
                          & \multirow{3}{*}{openfortivpn} & 07946c1  & 1                                                         \\
                          &                               & f755c99  & 1                                                         \\
                          &                               & 0007b2d  & 3                                                         \\ \cmidrule{2-4}
                          & ovs                           & 9da8b2f  & 1                                                         \\ \cmidrule{2-4}
                          & PHP                           & 7a4584d  & 1                                                         \\ \cmidrule{2-4}
                          & SPICE                         & ef9a8bf  & 1                                                         \\ \cmidrule{2-4}
                          & unbound                       & ffed368  & 1                                                         \\ \cmidrule{2-4}
                          &   \cellcolor[gray]{.8} httpd                         & \cellcolor[gray]{.8} 2.4.37   & \cellcolor[gray]{.8} 11                                                        \\  \cmidrule{2-4}
                          &  \cellcolor[gray]{.8}  curl                          & \cellcolor[gray]{.8} 7.63.0   & \cellcolor[gray]{.8} 7                                                         \\ 
\midrule[1pt]
\multirow{4}{*}{SQLite}   & \multirow{2}{*}{anope}        & 2a5e782  & 1                                                         \\ 
                          &                               & aeefe16  & 1                                                         \\ \cmidrule{2-4}
                          & darktable                     & 70820b1  & 1                                                         \\ \cmidrule{2-4}
                          & librdf                        & 5d074c1  & 1                                                         \\ 
\midrule[1pt]
\multirow{12}{*}{libpcap} & arp-scan                      & f013b45  & 1                                                         \\ \cmidrule{2-4}
                          & arping                        & b37fb24  & 1                                                         \\ \cmidrule{2-4}
                          & \multirow{3}{*}{ettercap}     & 89b5542  & 5                                                         \\ 
                          &                               & dfcabfc  & 2                                                         \\ 
                          &                               & 891a281  & 1                                                         \\ \cmidrule{2-4}
                          & freeradius                    & 57fbb95  & 1                                                         \\ \cmidrule{2-4}
                          & knock                         & 4b8ad4d  & 1                                                         \\ \cmidrule{2-4}
                          & libnet                        & 008c994  & 1                                                         \\ \cmidrule{2-4}
                          & ntop                          & 66f6f48  & 1                                                         \\ \cmidrule{2-4}
                          & \multirow{2}{*}{tcpdump}      & 39be365  & 1                                                         \\
                          &                               & 224b073  & 2                                                         \\ \cmidrule{2-4}
                          & wireshark                     & 51a99ca  & 1                                                         \\ 
\midrule[1pt]
\multirow{2}{*}{libxml2}  & \multirow{2}{*}{abiword}      & 80fee4c  & 2                                                         \\
                          &                               & ebcc445  & 4                                                         \\ 
\midrule[1pt]
\multicolumn{4}{c}{\boldmath $D_{Artif-UV}$}                                                                                     \\ 
\midrule[1pt]
\multirow{3}{*}{OpenSSL}  & OpenWrt                       & -        &   6                                                        \\
                          & Tomato-shibby                 & -        &   6                                                        \\
                          & AsusWrt                      & -        &    6                                                       \\ 
\midrule[1pt]
\multirow{3}{*}{SQLite}   & OpenWrt                       & -        &    6                                                       \\
                          & Tomato-shibby                 & -        &    6                                                       \\
                          & AsusWrt                      & -        &     5                                                      \\ 
\midrule[1pt]
\multirow{3}{*}{libpcap}  & OpenWrt                       & -        &   6                                                        \\
                          & Tomato-shibby                 & -        &    6                                                       \\
                          & AsusWrt                       & -        &    5                                                                                                         \\
\midrule[1pt]
\multirow{3}{*}{libxml2}  & OpenWrt                       & -        &    6                                                       \\
                          & Tomato-shibby                 & -        &    6                                                       \\
                          & AsusWrt                       & -        &    5                                                       \\
\bottomrule[1.5pt]
\end{tabular}
\end{table}

\begin{table}[]
\centering
\caption{API description extraction accuracy.}
\label{Table: API Description Extraction Accuracy}
\begin{tabular}{c|c|c}
\toprule[1.5pt]
\multirow{2}{*}{\bfseries{Tool}}    &  \multicolumn{2}{c}{\boldmath $D_{DESC_{test}}$}   \\ 
                                         
                                     & \bfseries{Accuracy}            & \bfseries{F1}        \\ \midrule[1pt]
\system                                    &92.41\% & 85.24\%  \\

\textit{Advance}~\cite{lv2020rtfm}       &85.07\% &74.37\%    \\ 
\textit{RCNN}~\cite{lai2015rcnn}   & 76.25\% & 61.50\%  \\
\textit{ALICS}~\cite{pandita2012inferring}   & 38.43\% & 29.02\% \\

\bottomrule[1.5pt]
\end{tabular}
\end{table}

\subsubsection{API Description Extraction Accuracy}
In this step, we evaluate the API description extraction accuracy of \system with two metrics: accuracy and F1 score. 
We also compare  \system with three off-the-shelf tools: \textit{Advance}~\cite{lv2020rtfm}, \textit{RCNN}~\cite{lai2015rcnn}, and \textit{ALICS}~\cite{pandita2012inferring}. \textit{Advance} is the first one that adopts semantic analysis to find API descriptions in TPC documents. It is important to mention that \textit{Advance} treats the API descriptions as API specifications directly. \textit{RCNN} is a popular classifier that can also be applied to sentiment analysis tasks. \textit{ALICS} leverages semantic templates with shallow parsing to recognize API descriptions.


We train \system from scratch on $D_{DESC_{train}}$ for 100 epochs with a batch size of 128. We set the learning rate to 1e-3 and the dropout rate to 0.25. We save the model when it achieves the best accuracy on $D_{DESC_{dev}}$ during the 100 epochs. We also set up \textit{Advance}, \textit{RCNN}, and \textit{ALICS} according to their instructions respectively.  Both \textit{Advance} and \textit{RCNN} are trained on $D_{DESC_{train}}$. As shown in Table~\ref{Table: API Description Extraction Accuracy}, our system achieves 92.41\% accuracy and 85.24\% F1 score on $D_{DESC_{test}}$, which are higher than the other three methods. The main reason for our excellent performance is that \system solves the coreference in TPC documents and makes good use of contextual information to identify the sentiment of APIs. Nevertheless, \textit{Advance} and \textit{RCNN} do not solve the coreference first and fail to fully leverage the contextual information. Besides, \textit{RCNN} does not leverage the attention mechanism, hindering its performance on the sentiment analysis task.
Besides, the template-based matching method adopted by \textit{ALICS} cannot correctly handle different TPC documents since they do not have a consistent format. In addition, we evaluate the generalization ability of \system on extracting API descriptions from new TPC documents. Specifically, \system achieves $89.05\%$ accuracy and $74.86\%$ F1 score on $D_{DESC_{new}}$, indicating its great ability to generalize when presented with new TPC documents.

We further explore the possible reasons for false positives and false negatives of \system as follows. First, our coreference resolution model cannot handle a part of coreferences in the TPC document, making it difficult to identify the exact sentiment of several API descriptions, causing false positives. Second, our sentiment analysis model cannot recognize some relevant API descriptions that do not exhibit a clearly expressed sentiment. For example, the description ``\textit{bindings are not cleared by the sqlite3\_reset() routine}" cannot be recognized by the model due to the lack of an apparent strong sentiment.

\begin{table}[]
\centering
\caption{API specification inference accuracy.}
\label{Table: API Specification Inference Accuracy}
\footnotesize
\begin{tabular}{l|c|c|c|c}
\toprule[1.5pt]
\multicolumn{1}{c|}{\multirow{2}{*}{\bfseries{Tool}}}    & \multicolumn{2}{c|}{\boldmath $D_{MRC_{dev}}$} & \multicolumn{2}{c}{\boldmath $D_{MRC_{test}}$}\\ 
                                     & \bfseries{EM}            & \bfseries{F1}  & \bfseries{EM}            & \bfseries{F1}          \\ \midrule[1pt]
\begin{tabular}[c]{@{}l@{}}\system (\textit{RoBERTa}) \\  + 
Distilled Documents\end{tabular}                            & 87.48\%       & 90.17\%       &86.52\% & 88.23\%   \\ \midrule[1pt]

\begin{tabular}[c]{@{}l@{}}\textit{RoBERTa}\\ + Original Documents\end{tabular}  & 79.34\%       & 81.06\%  & 76.91\% & 78.60\%   \\ \bottomrule[1.5pt]
\end{tabular}
\end{table}

\begin{table}[]
\centering
\small
\caption{Generalization ability and error propagation analysis on the MRC system of \system.}
\label{Table: Generalization ability and error propagation analysis}
\begin{tabular}{l|c|c}
\toprule[1.5pt]
\multicolumn{1}{c|}{\multirow{2}{*}{\bfseries{Tool}}}    &  \multicolumn{2}{c}{\boldmath $D_{MRC_{new}}$}   \\ 
                                         
                                     & \bfseries{EM}            & \bfseries{F1}        \\ \midrule[1pt]
\begin{tabular}[c]{@{}l@{}}\system (\textit{RoBERTa}) \\  + Manually Distilled Documents\end{tabular}                                  &88.19\% & 90.85\%  \\  \midrule[1pt]
\begin{tabular}[c]{@{}l@{}}\system (\textit{RoBERTa}) \\  + Automatically Distilled Documents\end{tabular}                              &84.40\% & 85.89\%    \\ 

\bottomrule[1.5pt]
\end{tabular}
\end{table}

\vspace{-0.2cm}
\subsubsection{API Specification Inference Accuracy}
In this step, we evaluate the API specification inference accuracy of \system with two metrics: Exact Match (EM) and F1 score. EM represents the percentages of predicted answers that match any of the ground-truth answers. 
EM and F1 score are two major metrics used in SQuAD\cite{DBLP:conf/emnlp/RajpurkarZLL16,DBLP:conf/acl/RajpurkarJL18}, which is a well-known MRC dataset, to evaluate the performance of MRC systems. In this paper, we also adopt these two metrics to evaluate \system. 

We set the learning rate to 1e-5 and train \system (\textit{RoBERTa}) for 500K steps with a batch size of 256 sequences on $D_{MRC_{train}}$. As shown in Table~\ref{Table: API Specification Inference Accuracy}, the EM and F1 score of \system are 87.48\% and 90.17\% respectively on the development set, and 86.52\% and 88.23\% respectively on the testing set. To further evaluate the importance of document distillation, we conduct an ablation experiment by directly comparing \system with \textit{RoBERTa}. 
More specifically, \textit{RoBERTa} is trained on $D_{MRC_{train}}$ with the same parameters as \system but without eliminating irrelevant API descriptions. Since our MRC system is developed from \textit{RoBERTa} as well, this ablation experiment can be regarded as we perform \textit{RoBERTa} on distilled documents and original documents, respectively. The distilled documents are obtained based on our  document distillation model.
The results show that \system outperforms \textit{RoBERTa} on both $D_{MRC_{dev}}$ and $D_{MRC_{test}}$, highlighting the significance of document distillation. During the analysis, we notice that \textit{RoBERTa} extracts incorrect answers from irrelevant API descriptions, resulting in false positives. 

Besides, although the MRC system of \system is developed from \textit{RoBERTa}, which has already demonstrated great generalization capabilities, we still evaluate its generalization ability on $D_{MRC_{new}}$. In addition, we analyze the error propagation between the API description extraction module and the API specification inference module by performing \system on $D_{MRC_{new}}$ with manually distilled and automatically distilled documents, respectively.  
Notably, as shown in Table~\ref{Table: Generalization ability and error propagation analysis}, \system achieves $84.40\%$ EM and $85.89\%$ F1 score on $D_{MRC_{new}}$ with automatically distilled documents, highlighting its great capacity to generalize when faced with new TPC documents. Moreover, both the EM and F1 score on manually distilled documents are higher than those on automatically distilled documents, demonstrating error propagation between the two modules. The false positives and false negatives caused by the API description extraction module have an influence of about 5\% on the result of the API specification inference module.

We further analyze the possible reasons for false positives and false negatives caused by the MRC system of \system as follows. First, \system wrongly extracts answers from TPC documents for unanswerable questions, leading to false positives. Addressing the unanswerable questions is still an open research question for the MRC task.
Second, complex or ambiguous API descriptions cannot be properly handled by \system, which may also cause false positives and false negatives.  For instance, the description ``\textit{the return values of the SSL*\_ctrl() functions depend on the command supplied via the cmd parameter}'' lacks clarity in stating the return values of \texttt{SSL*\_ctrl()}, and this information is also not provided in other related descriptions. Besides, the function name \texttt{SSL*\_ctrl()} has a wildcard character and is not a standard one, making it challenging for our MRC system to comprehend its meaning.




\begin{table}[]
\centering
\caption{Programming expression generation accuracy.}
\label{Table: Programming Expression Generation Accuracy}
\begin{tabular}{c|c|c}
\toprule[1.5pt]
\multirow{2}{*}{\bfseries{Tool}}    &  \multicolumn{2}{c}{\boldmath $D_{PE}$}\\ 
                                     & \bfseries{Recall}            & \bfseries{FNR}          \\ \midrule[1pt]
\system                                    &82.17\% & 17.83\%   \\

\textit{Jdoctor}~\cite{blasi2018jdoctor}   & 30.50\% & 69.50\%  \\
\textit{DRONE}~\cite{zhou2018api}   & 50.33\% &    49.67\%\\
\textit{Advance}~\cite{lv2020rtfm}       &72.50\% &27.50\%    \\ 
\bottomrule[1.5pt]
\end{tabular}
\vspace{-2em}
\end{table}

\vspace{-0.2cm}
\subsubsection{Programming Expression Generation Accuracy} In this step, we evaluate the programming expression generation accuracy of \system with two metrics: recall and false negative rate.  We also compare \system with three state-of-the-arts: \textit{Jdoctor}~\cite{blasi2018jdoctor},  \textit{DRONE}~\cite{zhou2018api}, and \textit{Advance}. \textit{Jdoctor} leverages the pattern matching and lexical matching to translate specifications into expressions. \textit{DRONE} uses more than 60 heuristics to translate Java API specifications to expressions.
We follow the instructions declared in their papers to set up them and perform them on $D_{PE}$.


As shown in Table~\ref{Table: Programming Expression Generation Accuracy}, \system achieves a recall of 82.17\% and a false negative rate of 17.83\%, which is the highest recall and the lowest false positive rate among all methods. The improvement from \textit{Advance} to \system shows the effectiveness of extracting more precise API specifications before generating programming expressions. As for another two methods, first, \textit{Jdoctor} performs well in translating API specifications with simple arithmetic and logical operations but fails to deal with complicated API specifications. Second, for \textit{DRONE}, we find its more than 60 heuristics are well suitable for Java API documents but do not perform well on C API documents.

Our further analysis shows that the false negatives of \system are mainly due to the incorrect API specifications inferred from several API descriptions, leading us to generate the erroneous programming expressions. Besides, though we maintain lots of patterns to map phrases into programming expressions, they cannot cover several uncommon phrases, leading to false negatives. For instance, we lack a pattern to match the uncommon phrase ``\textit{must be an index},'' which appears only within the following API description: ``\textit{the second argument must be an index into the aConstraint[] array}.''



\begin{table}[]
\caption{Usage violation detection accuracy.}
\label{Table: Usage Violation Detection Accuracy}
\centering
\footnotesize
\setlength\tabcolsep{2pt}
\label{Table: Detection Accuracy}
\begin{tabular}{c|c|c|c|c|c|c}
\toprule[1.5pt]
\multicolumn{1}{c|}{\bfseries{\multirow{2}{*}{Performance}}} & \multicolumn{3}{c|}{{\system}} & \multirow{2}{*}{\bfseries{\textit{Advance}}} & \multirow{2}{*}{\bfseries{\textit{APISAN}}} & \multirow{2}{*}{\bfseries{\textit{APEx}}} \\ 
\multicolumn{1}{c|}{}                  & \bfseries{x86}     & \bfseries{ARM}     & \bfseries{MIPS}    &                                   &                                  &                                \\ \midrule[1pt]
\multicolumn{7}{c}{\boldmath $D_{Real-UV}$}                                                                                                                                           \\ \midrule[1pt]
Precision                              &     72.84\%    &  74.70\%       &   77.03\%      &     80.72\%                              &     17.31\%                             &    23.07\%                            \\
Recall                                 &   76.62\%      &  80.52\%       &  74.03\%       &         87.01\%                         &      11.69\%                            &   7.79\%                             \\

\midrule[1pt]
\multicolumn{7}{c}{\boldmath $D_{Artif-UV}$ }                                                                                                                                           \\ \midrule[1pt]
Precision                              &    68.92\%    &  74.32\%       &   76.47\%      &     77.33\%                              &     25.49\%                             &    34.62\%                            \\
Recall                                 &   73.91\%      &  79.71\%       &   75.36\%      &         84.06\%                         &      18.84\%                            &   13.04\%                             \\ \bottomrule[1.5pt]
\end{tabular}
\vspace{-1pt}
\end{table}

\vspace{-0.2cm}
\subsubsection{Usage Violation Detection Accuracy}
In this step, we leverage two metrics, precision and recall, to evaluate the usage violation detection accuracy of \system.  Since we are the first to address the TPC usage violation problem in binary IoT firmware, we cannot conduct the exact same comparison with previous works. We finally choose to compare with three source-level API misuse detectors:  \textit{Advance}, \textit{APISAN}~\cite{DBLP:conf/uss/YunMSJKN16}, and \textit{APEx}~\cite{DBLP:conf/kbse/KangRJ16}. \textit{APISAN} first infers API specifications by analyzing the source code of various usage examples of the API in C programs, and then implements a series of checkers for different misuse cases. \textit{APEx} focuses on API error specifications. It first analyzes the error-handling code in multiple API call sites to get error constraints, and then performs under-constrained symbolic execution to find the misuses. Besides, since we only compare with fully automated API misuse detectors, we do not compare with \textit{ARBITRAR}~\cite{li2021arbitrar}, which requires human interaction when executing the system. In summary, three API misuse detectors perform on the source-level while \system performs on the binary-level. 
We perform \system on $D_{Real-UV}$ and $D_{Artfi-UV}$ respectively. The programs in these two datasets are compiled into three different architectures (x86, MIPS, and ARM) along with two different optimization levels (O0 and O3) under GCC 5.4.0. 

For \system on $D_{Real-UV}$, as shown in Table~\ref{Table: Usage Violation Detection Accuracy}, it achieves 72.84\% precision,  76.62\% recall on x86, 74.70\% precision and 80.52\% recall on ARM, and 77.03\% precision and 74.03\% recall on MIPS. For \system on $D_{Artfif-UV}$, it achieves 68.92\% precision and 73.91\% recall on x86, 74.32\% precision and 79.71\% recall on ARM, and 76.47\% precision and 75.36\% recall on MIPS. Though $D_{Real-UV}$ and $D_{Artfi-UV}$ have the ground-truth, we still manually check the reported results to avoid the potential corner cases. The results show that \system has great performance across different architectures. Besides, we notice that though \system is performed on binary-level, it still has much higher precision and recall than \textit{APISAN} and \textit{APEx}, and is very close to the performance of \textit{Advance}. This is because we infer precise API specifications from TPC documents and design a rule-driven analysis engine that can handle complex logical inference.
We further analyze the false positives and false negatives of these API misuse detectors. For \textit{APISAN} and \textit{APEx}, first, we find both of them cannot detect deprecated APIs. Second, most of their inferred API specifications are incomplete or incorrect, leading to false positives and false negatives. The main reason is that \textit{APISAN} and \textit{APEx} heavily rely on analyzing the source-level C programs, which may contain the API usage examples, to infer the API specifications. Nevertheless, the C programs may not use all APIs in a TPC and even cannot cover all possible usage examples for an API, resulting in incomplete or incorrect API specification inference. For \textit{Advance}, though its inferred API specifications are more accurate than \textit{APISAN} and \textit{APEx}, the incorrect verification code and the limited performance of \textit{CodeQL} cause the false positives and false negatives. For instance, \textit{CodeQL} fails to perform a dataflow analysis in some complicated cases, leading to false positives.

In addition, to find the causes of false positives and false negatives of \system, we further explore them as follows. First, incorrect and imprecise API specifications are the main reason for the false positives and false negatives. Currently, we cannot extract the API specifications without errors. Second, incorrect programming expressions may also lead to multiple false positives and false negatives. 

\vspace{-0.3cm}
\section{Large-scale Analysis on IoT Firmware}
To obtain an in-depth status quo understanding of the TPC usage violation problem in IoT firmware, we further leverage \system to conduct a large-scale analysis on $4,545$ firmware images. In this section, we aim to answer the following research questions.

\noindent$\bullet$ \textbf{RQ1:} Which are the most prevalent TPC usage violations in IoT firmware?

\noindent$\bullet$ \textbf{RQ2:} What are the practical impacts of TPC usage violations on IoT firmware?

\vspace{-0.4cm}
\subsection{Experimental Setup}
Before conducting the analysis, we need to first construct a large-scale firmware dataset. To achieve this goal, we obtain $34,136$ firmware images from \textit{FirmSec}~\cite{zhao2022large}, containing $35$ different kinds of firmware from hundreds of vendors. These firmware images use a total of 584 TPCs. Since we study four TPCs for concept validation, we do not perform the analysis on all firmware images. Therefore, we first use the firmware processing module of \system to identify the firmware images that employ the four TPCs we studied. Additionally, we cross-check our results by using the original analysis results obtained from the authors of \textit{FirmSec}. As shown in Table~\ref{Table: Firmware Dataset Composition}, we find $4,545$ firmware images from 9 vendors using at least one of the four TPCs. 
More specifically, $4,126$ firmware images adopt OpenSSL, $2,253$ firmware images contain SQLite, $3,820$ firmware images use libpcap, and $1,229$ firmware images utilize libxml2.

\begin{table}[]
\caption{Firmware dataset composition.}
\label{Table: Firmware Dataset Composition}
\renewcommand\arraystretch{0.7}
\tiny
\centering
\begin{tabular}{p{1cm}<{\centering}|p{1.3cm}<{\centering}|c|c|c|c|c}
\toprule[1.5pt]
\multirow{2}{*}{\bfseries{Vendor}}   & \multirow{2}{*}{\bfseries{Category}} & \multirow{2}{*}{\bfseries{\# Firmware}} & \multicolumn{4}{c}{\bfseries{TPC}}              \\ 
                          &                           &                              & \bfseries{OpenSSL} & \bfseries{SQLite} & \bfseries{libpcap} & \bfseries{libxml2} \\ \midrule[1pt]
\multirow{4}{*}{D-Link}   & IP Camera                 & 156                          & 156     & 44     & 151     & 0       \\ \cmidrule{2-7} 
                          & Router                    & 437                          & 424     & 224    & 254     & 0       \\ \cmidrule{2-7}
                          & Switch                    & 49                           & 49      & 0      & 49      & 0       \\ \cmidrule{2-7}
                          & Smart Home                & 21                           & 21      & 16     & 12      & 0       \\ \midrule[1pt]
\multirow{2}{*}{TP-Link}  & IP Camera                 & 268                          & 268     & 87     & 226     & 0       \\ \cmidrule{2-7}
                          & Router                    & 757                          & 725     & 297    & 711     & 0       \\ \midrule[1pt]
\multirow{3}{*}{TRENDnet} & IP Camera                 & 225                          & 155     & 40     & 170     & 0       \\ \cmidrule{2-7}
                          & Router                    & 255                          & 216     & 64     & 163     & 0       \\ \cmidrule{2-7}
                          & Switch                    & 112                          & 112     & 10     & 112     & 0       \\ \midrule[1pt]
DAHUA                     & IP Camera                 & 207                          & 88      & 10     & 204     & 0       \\ \midrule[1pt]
Xiongmai                  & IP Camera                 & 105                          & 38      & 4      & 104     & 0       \\  \midrule[1pt]
Fastcom                   & Router                    & 103                          & 48      & 16     & 98      & 0       \\ \midrule[1pt]
Xiaomi                    & Router                    & 20                           & 20      & 20     & 20      & 0       \\ \midrule[1pt]
TSmart                    & Smart Home                & 344                          & 336     & 61     & 263     & 0       \\ \midrule[1pt]
OpenWrt                   & Router                    & 1,486                        & 1,470   & 1,360  & 1,283   & 1,229   \\ \midrule[1pt]
\multicolumn{2}{c|}{Overall}   & 4,545                        & 4,126   & 2,253  & 3,820   & 1,229    \\ \bottomrule[1.5pt]
\end{tabular}
\end{table}

\subsection{Usage Violation Distribution}
This subsection answers RQ1. As shown in Table~\ref{Table: Usage Violation Distribution}, we present the large-scale usage violation detection results on our firmware dataset. We find the firmware dataset consists of consecutive firmware sets, involving at least 937 firmware images. Each consecutive firmware set includes a series of historical firmware images belonging to the same device. Therefore, it is possible that the same usage violation exists in multiple historical firmware versions associated with the same device. In our analysis, we treat each historical firmware image as a distinct entity. The same violation that exists in different historical firmware will be counted cumulatively. Besides, to get more accurate results of the deprecated API violation, we maintain version-specific deprecated API lists for each TPC in the large-scale evaluation. According to the results, we have the following findings. 

\textbf{First}, TPC usage violations are widespread in IoT firmware. More specifically, we detect $27,621$ usage violations of the four TPCs in the 4,545 firmware images, including 11,348 deprecated API violations, 8,304 causality violations, 5,891 return value violations, and 2,078 argument violations.
\textbf{Second}, the deprecated API violation takes the majority of the overall usage violations, accounting for $41\%$. \textbf{Third}, the causality violation and return value violation are also widespread in IoT firmware, accounting for $30\%$ and $21\%$, respectively. For the causality violation, most usage violations are due to lacking a call for some APIs. \textbf{Finally}, more than $40\%$ usage violations are inherited from other TPCs adopted by firmware. For instance, the latest version (2.22.19) of Xiaomi Router Mini is still using cURL 7.33.0, which contains a number of OpenSSL usage violations. Besides, the four TPCs also have some internal usage violations. For example, we find that the OpenSSL used in many TP-Link does not check the return value of \texttt{BN\_CTX\_get}, which may cause the Denial-of-Service attack.

We have reported all the usage violations to the corresponding vendors. Up to now, D-Link, TP-Link, Xiaomi, and TSmart have responded to us and 206 usage violations have been confirmed by vendors as vulnerabilities, and seven of them have been assigned CVE IDs with high severity. In addition, at least 117 of the 206 confirmed violations have been fixed.




\begin{table}[]
\centering
\caption{Usage violation distribution.}
\label{Table: Usage Violation Distribution}
\setlength\tabcolsep{2pt}
\footnotesize
\begin{tabular}{c|c|c|c|c}
\toprule[1.5pt]
\bfseries{TPC}      & \begin{tabular}[c]{@{}c@{}}\bfseries{\# Deprecated API}\\ \bfseries{Violation}\end{tabular} & \begin{tabular}[c]{@{}c@{}} \bfseries{\#Causality}\\ \bfseries{Violation}\end{tabular} & \begin{tabular}[c]{@{}c@{}}\bfseries{\# Return Value}\\ \bfseries{Violation}\end{tabular} & \begin{tabular}[c]{@{}c@{}}\bfseries{\# Argument}\\ \bfseries{Violation}\end{tabular} \\ \midrule[1pt]
OpenSSL        &      4,831                                                                &   3,679                                                             &  3,521                                                                   &   1,073                                                          \\ 
SQLite             &  2,740                                                                    &     1,996                                                            &      931                                                             &  112                                                               \\
libpcap               &    3,359                                                                   &     2,515                                                             &        1,364                                                             &     857                                                            \\ 
libxml2               &    418                                                                   &    114                                                              &   75                                                                  &  36  \\ \midrule[1pt]
Overall               &    11,348                                                                   &    8,304                                                              &   5,891                                                                  &  2,078  \\ \bottomrule[1.5pt]                                                            
\end{tabular}
\end{table}


\vspace{-0.2cm}
\subsection{Practical Impacts}
This subsection answers RQ2. Though we have identified many usage violations in IoT firmware, we still lack an understanding of the practical impacts of them. Therefore, we conduct further analysis to explore the potential consequences of these usage violations and gain deeper insights into their practical impacts. We manually analyze hundreds of usage violations and collect feedback from vendors about our reported usage violations. Based on the analysis results, we classify the possible impacts of TPC usage violations into three categories: security vulnerabilities, ordinary bugs, and no impacts. \textbf{First}, our analysis identifies a set of usage violations that could be exploited to perform attacks, e.g., the Man-In-The-Middle attack, and we regard these types of violations as security vulnerabilities. To provide real-world examples of the attacks that could be carried out using these violations, we conduct two case studies, which are discussed in the following subsections. \textbf{Second}, our analysis reveals that many usage violations are ordinary bugs that may result in the malfunctioning of firmware but cannot be leveraged for attacks. One such example is found in certain TP-Link firmware images that exhibit an incorrect verification of the return value of  \texttt{SSL\_write()} as $<0$ instead of $<=0$, causing the usage violation.  TP-Link regards this type of violation as a conventional bug rather than a vulnerability as it merely causes functional errors without posing any threat to potential attacks.
\textbf{Finally}, we discover that some usage violations have no impact on firmware. For instance, we detect some firmware images that do not call \texttt{SSL\_get\_error()} when \texttt{SSL\_do\_handshake()} returns 0, though it is a requirement stated in the OpenSSL document. Nevertheless, since \texttt{SSL\_get\_error()} only returns the error code for diagnosis purposes, this particular violation does not impact the firmware.

\vspace{-0.2cm}
\subsubsection{Denial-of-Service Attack} Based on our further analysis, we find some firmware images from TSmart do not check the return value of many APIs from OpenSSL, including  \texttt{SSL\_renegotiate}, \texttt{SSL\_do\_handshake}, \texttt{SSL\_peek}, and \texttt{SSL\_set\_session\_id\_context}. 
Lacking a check for the return value of the above APIs will result in serious consequences, e.g., the Denial-of-Service attack. For instance, \texttt{SSL\_renegotiate} is an essential API to handle the SSL renegotiation process between the client and the server. It has two return values that indicate the success or error status respectively. The SSL renegotiation process usually consumes many computing resources since it involves complicated calculation. If the server does not check the return value of \texttt{SSL\_renegotiate}, the server will mistakenly enter the next step, e.g., \texttt{SSL\_do\_handshake}, leading to more waste of resources. Therefore, if some low-power devices, e.g., IoT devices, allow insecure renegotiation and fail to impose restrictions on renegotiation requests, an attacker could initiate many renegotiation requests to exhaust their computing resources, causing a Denial-of-Service attack.

To exploit this vulnerability, we mainly employ emulating-based dynamic analysis. We here adopt \textit{FIRMADYNE}~\cite{chen2016towards} to conduct dynamic analysis on the firmware. \textit{FIRMADYNE} supports emulating the Linux-based firmware on the desktop. It overcomes the general challenges in emulating firmware, such as the presence of hardware-specific peripherals. After emulating the firmware, we create many malicious clients to send renegotiation requests to the firmware. Finally,  we successfully perform the Denial-of-Service attack on the vulnerable firmware.

\begin{figure}[H]
\centering
\includegraphics[scale=0.38]{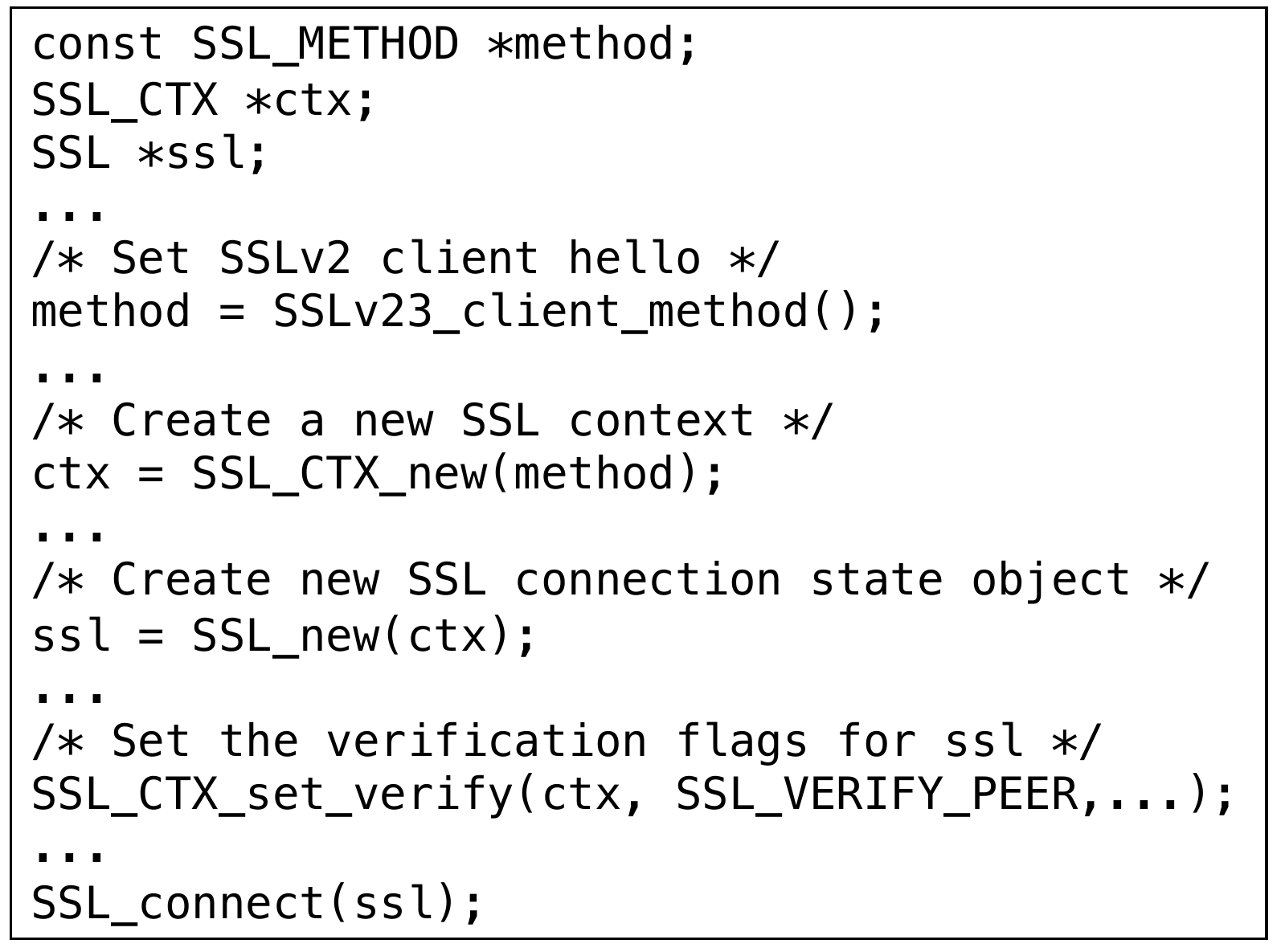}
\caption{SSL certificate validation process.}
\label{Figure:  SSL certificate validation process}
\vspace{-0.45cm}
\end{figure}

\vspace{-0.2cm}
\subsubsection{Man-In-The-Middle Attack}
Typically, firmware adopts OpenSSL to perform certificate validation during SSL/TLS handshake. Figure~\ref{Figure:  SSL certificate validation process} presents a standard SSL certificate validation process. Nevertheless, we find
a set of firmware images from D-Link lack a call to the function \texttt{SSL\_CTX\_set\_verify}, which is an essential function for SSL certificate validation.
Without calling \texttt{SSL\_CTX\_set\_verify}, attackers could forge a certificate to deceive devices and then capture sensitive information or compromise devices, leading to Man-In-The-Middle attacks. 


To exploit this vulnerability, we also  adopt \textit{FIRMADYNE} to conduct dynamic analysis on the firmware.  After emulating the  firmware, we create a host with a forged certificate in the same local area network. We leverage ARP spoofing, also known as ARP cache poisoning, to attack the emulated firmware and the gateway, causing any traffic meant for the targeted host
to be sent to our host instead. We then redirect the traffic to the target host and successfully establish the connection with the emulated firmware and the target host separately. Finally, we successfully obtain the plaintext transferred between the firmware and the server, which includes sensitive information, e.g., password.

We have contacted D-Link to report the vulnerability, and they tell us they have fixed this vulnerability in the past~\cite{Dlinkvul}. It is reasonable since most firmware images in the dataset were collected by Zhao et al.~\cite{zhao2022large} several years ago. Nevertheless, we find this vulnerability still poses a threat to a significant number of D-Link devices.  We leverage \textit{Shodan}~\cite{shodan} to search affected D-Link devices and discover more than $10,000$ vulnerable devices are still publicly available on the Internet.

\vspace{-0.4cm}
\section{Discussion}
\noindent\textbf{Ethics.} 
We pay special attention to the potential ethical issues in this work. First, we obtain all the used firmware from legitimate sources. For the firmware images in $D_{Artif-UV}$,
we collect them from the corresponding official websites. For the firmware images used in the large-scale analysis, we collect them from \textit{FirmSec} and strictly follow the research-only agreement.
Second, we have done responsible disclosure to report all the detected usage violations to vendors. 

\noindent\textbf{Limitations and future work.}  
Though \system achieves a great performance in detecting the TPC usage violation in IoT firmware, it still has several limitations for future research as follows. 
\textbf{First}, \system has false positives and false negatives. In our experience, the false positives and false negatives are mainly introduced by incorrect API specifications and incorrect rule-driven analysis. Though our NLP model achieves high accuracy in API specification inference, it still cannot process some complicated cases. Meanwhile, our rule-driven  analysis mainly targets four kinds of usage violations, which may not cover all possible cases. Besides, some usage violations are extremely complicated and cannot be appropriately handled by \system. In the future, we will improve the performance of our NLP model and enhance \system to cover more complicated cases. \textbf{Second}, \system currently is limited to analyzing firmware of three architectures: x86, ARM, and MIPS. In the future, we will extend the analysis to more architectures, e.g., RISC-V, by adding new disassemblers on top of our binary-to-Datalog encoding tool.
\textbf{Third}, \system does not consider code obfuscation when analyzing firmware. In this paper, we leverage \textit{binwalk} to unpack and extract firmware, but it only supports analyzing the firmware without code obfuscation. Unpacking firmware with code obfuscation is still an open research question and is out of our research scope. Our paper aims to fill the gap in mapping the high-level specifications
from TPC documents to the binary-level violation analysis. Besides, based on our analysis, we find most firmware images are not obfuscated. Therefore, we do not consider code obfuscation in this paper.  
It will be interesting future work to enhance the ability of \system to analyze obfuscated firmware.

\section{Related work}

\textbf{Firmware analysis.} 
Many works have studied the security of IoT firmware through static or dynamic analysis. Currently, machine learning techniques are widely adopted by static analysis methods. For instance, Xu et al.~\cite{xu2017neural} utilized the neural network to detect similar vulnerable codes in different binaries. Ding et al.~\cite{ding2019asm2vec} adopted learned representation to conduct assembly clone detection in binaries. Zhao et al.~\cite{zhao2022large} conducted the first large-scale analysis of the vulnerabilities introduced by TPCs in IoT firmware based on a customized neural network. Dynamic analysis methods usually perform testing on the emulated firmware.
For example, Chen et al.~\cite{chen2016towards} proposed \textit{FIRMADYNE} to emulate IoT firmware on the desktop and conducted a dynamic analysis to find the vulnerabilities in IoT firmware. Feng et al. designed \textit{$P^{2}IM$}~\cite{feng2020P2IM}, a scalable and hardware-independent method, to conduct MCU firmware fuzzing. In addition, a set of works have explored firmware re-hosting techniques, which are essential for dynamic firmware analysis. Clements et al. designed \textit{HALucinator}~\cite{clements2020halucinator}, a high-level emulation system that can re-host firmware through abstraction layer modeling. Scharnowski et al. implemented \textit{FuzzWare}~\cite{scharnowski2022fuzzware} which is a highly efficient monolithic firmware fuzzing system that employs a novel re-hosting technique to avoid path elimination. Nevertheless, applying these methods in our task is not trivial since they are not designed to detect TPC usage violations in IoT firmware.


\noindent\textbf{API misuse detection.} 
A series of works have been proposed to detect API misuses at the source-level. These works can be divided into three categories according to their different API specification inference strategies. The first kind of work obtains API specifications by analyzing many API usage examples. For example, Yun et al.~\cite{DBLP:conf/uss/YunMSJKN16} inferred correct API
usage by analyzing the symbolic contexts of many API usage examples. Kang et al.~\cite{DBLP:conf/kbse/KangRJ16} obtained error specifications for API functions by analyzing a wealth of corresponding usage examples across multiple C programs. The second kind of methods is to obtain API specifications from generated API usage examples. Wen et al.~\cite{wen2019mutation} adopted mutation analysis to generate API usage examples from the correct ones. Nevertheless, the above two methods usually result in low precision since obtaining comprehensive correct usage examples of an API is hard. The third kind of methods is to obtain API specifications from the document. For instance, Lv et al.~\cite{lv2020rtfm} utilized multiple NLP techniques to infer API specifications from the document.
Except for the above three strategies, more recently, Li et al.~\cite{li2021arbitrar} designed a novel method using active learning to interact with users to detect API misuses. However, this method requires multiple rounds of user interaction which cannot be performed automatically. In this paper, we infer API specifications from TPC documents, as this approach has higher accuracy than other strategies. Compared with similar works that adopt the same strategy, our method heuristically solves the coreference in TPC documents and makes good use of contextual information to infer API specifications, leading to high precision and recall. Moreover, though the above source-level detectors can obtain API specifications, they cannot be applied to our task. This is because their static analysis tools, e.g., \textit{CodeQL}, are designed for the source-level analysis, and IoT firmware images are mostly closed-source binaries.



\noindent\textbf{Datalog-based static analysis.}
Currently, several works have applied Datalog to static analysis. 
Grech et al.~\cite{grech2019datalog} proposed \textit{Gigahorse} to decompile Ethereum smart contracts by using Datalog. Besides, the development of \textit{Souffl{\'e}}~\cite{jordan2016souffle}, a Datalog variant, also inspires many tools to leverage Datalog for static analysis. For instance, Flores-Montoya et al.~\cite{flores2020datalog} designed \textit{Ddisasm} to disassemble the binary based on \textit{Souffl{\'e}}. 
These works prove the efficiency of Datalog in static analysis and inspire us to adopt Datalog to explore the TPC usage violation problem in IoT firmware.

\vspace{-0.4cm}
\section{Conclusion}
In this paper, we present \system, the first automated and practical
system to detect TPC usage violations in binary IoT firmware, which fills the gap in mapping the high-level specifications
from TPC documents to the binary-level violation analysis.
\system achieves more than 70\% precision and recall in
detecting TPC usage violations in binary IoT firmware, which has an excellent performance improvement even compared to state-of-the-art source-level works. Moreover, we conduct the first large-scale analysis of the TPC usage violation problem in IoT firmware.  We identify 27,621 usage violations caused by four popular TPCs in 4,545 firmware images. Up to now, 206 usage violations have been confirmed by vendors as vulnerabilities, and seven of them have been assigned CVE IDs with high severity.  

\vspace{-0.3cm}
\section*{Acknowledgments}

We sincerely appreciate the guidance from the shepherd. 
We would also like to thank the anonymous reviewers for their valuable comments to improve our paper. This work was partly supported by NSFC under No. U1936215, the State Key Laboratory of Computer Architecture (ICT, CAS) under Grant No. CARCHA202001, the Fundamental Research Funds for the Central Universities (Zhejiang University NGICS Platform), NSF under No. 1943100 and 1920462, and Meta Faculty Award.


\bibliographystyle{plain}
\bibliography{Ref}


%
\appendix

\end{document}